\documentclass[12pt]{article}

\newcommand{\be}{\begin{equation}}
\newcommand{\ee}{\end{equation}}
\newcommand{\ba}{\begin{eqnarray}}
\newcommand{\ea}{\end{eqnarray}}
\newcommand{\la}{\lambda}

\newcommand{\tr}{\rm tr}

\begin{document}
\hoffset=-.4truein\voffset=-0.5truein
\setlength{\textheight}{8.5 in}
\setcounter{equation}{0}
\begin{titlepage}
\begin{center}
\hfill August 9, 2011\\
\vskip 0.6 in
{\large \bf  { On  an Airy matrix  model with a logarithmic potential}}
\vskip .6 in
\begin{center}
 {{\bf E. Br\'ezin$^{a)}$}\\{\it and}\\{\bf S. Hikami$^{b),c)}$}}
\end{center}
\vskip 5mm
\begin{center}

{$^{a)}$ Laboratoire de Physique
Th\'eorique, Ecole Normale Sup\'erieure}\\ {24 rue Lhomond 75231, Paris
Cedex
05, France. e-mail: brezin@lpt.ens.fr{\footnote{\it
Unit\'e Mixte de Recherche 8549 du Centre National de la
Recherche Scientifique et de l'\'Ecole Normale Sup\'erieure.
} }}
\\
{$^{b)}$ Department of Basic Sciences,
} {University of Tokyo, Tokyo 153, Japan. }\\
{$^{c)}$ Okinawa Institute of Science and Technology,\\Tancha,Onna-son,Okinawa 904-0412,Japan,
e-mail:hikami@oist.jp}

\end{center}     
\vskip 3mm         

{\bf Abstract}                  
\end{center}
The Kontsevich-Penner model, an  Airy matrix model with a logarithmic potential, 
 may be  derived  from
a simple Gaussian two-matrix model through a duality. In this dual version the Fourier 
transforms of the n-point correlation functions can be computed in closed form. Using 
Virasoro constraints, we find that 
 in addition to the  parameters  $t_n$, which appears in the KdV hierarchies, one needs to introduce  here half-integer indices $t_{n/2}$ .
 The free energy as a function of those parameters may be obtained from these  Virasoro constraints.
The large N limit  follows from the solution to 
an integral equation.
This leads to explicit computations for a number of   topological invariants.

\end{titlepage}
\vskip 3mm

\section{Introduction}
\setcounter{equation}{0}
\renewcommand{\theequation}{1.\arabic{equation}}
  In some recent articles \cite{BH1,BH2,BH3,BH4,BH4a}, 
we have discussed a  relationship between the Airy matrix model (Kontsevich model)
and a Gaussian random matrix theory with an external source. 
The free energy for  higher Airy matrix models  (degree more than three) is the generating function for the intersection numbers
of the moduli space of $p$-spin  curves \cite{Witten1,Witten2,Kontsevich}.
 We have shown that the Fourier transform 
of the n-point correlation function of the Gaussian random matrix is also a generating
function of these intersection numbers with n-marked points \cite{BH3,BH4}. 
The reason for this remarkable agreement  is a duality for  expectation values of 
 characteristic polynomials \cite{BH2}.

It is interesting to extend this duality to the case of the $c=1$ \cite{B} matrix model, i.e. models dealing with a 1D chain of coupled matrices, and to understand the meaning of 
the topological invariants. 
In  \cite{BH4}, 
we have shown that the  time dependent random matrix theory, is equivalent to a
two-matrix model. These matrices are $N\times N$ and one considers correlation functions involving $k_1$ points for the first matrix, and $k_2$ for the second. After duality, we have shown that the edge   scaling limit is an  Airy matrix model with a logarithmic potential
\be\label{intro1}
Z = \int dB e^{-\frac{1}{3} {\rm tr} B^3 + k_2 {\rm tr}{\rm log} B + {\rm tr} B \Lambda}.
\ee
where $B$ is a $k_1\times k_1$ Hermitian matrix, and  $k_2$ is  the number of 
characteristic polynomials  in the two-matrix model \cite{BH4}. In  section 2, we recall
the derivation of (\ref{intro1}).

This Airy matrix model with a logarithmic potential, the so called Kontsevich-Penner model, 
 and its generalization to higher Airy
matrix model (degree larger than three) have been discussed in the literature.
For  the  general $(p+1)$-valent vertex model 
 $V(B) =  B^{p+1}/(p+1)$, the partition function
\be\label{intro2}
Z= \int dB e^{{\tr}V(B) + k {\tr} {\rm log} B +  {\tr} B \Lambda}
\ee
 was considered by Mironov et al \cite{Mironov}, in  particular for small $\Lambda$,  through
character expansions.
The parameter $\Lambda$ separates (i) the strong coupling  and (ii) the weak coupling  regions.
The transition  between (i) and (ii) is similar to the Br\'ezin-Gross-Witten transition 
\cite{BG,GW}
in the unitary matrix model. 
Indeed, it can be shown that the partition function $Z$ in (\ref{intro2}) with $p=-2$ is equivalent to the unitary
matrix model (see Appendix B). For $p=1$, it has been studied in 
\cite{Chekhov,Kostov}.

 The  model (\ref{intro2}) has been considered with
 polynomial vertices
 \be
V(B) = \sum \bar t_n B^n.
\ee
as a generating function of tachyon amplitudes in the c=1 string theory  \cite{Mukhi1,Mukhi2,Ambjorn}.
The existence of a  two dimensional black hole \cite{Susskind,Wadia,Mukhi3} has been discussed
 in related matrix models. 
We also note that
 some time evolution problems, such as crystal growth or non-intersecting random walks, share interesting universal  features
described by  the edge singularity of a random matrix theory
\cite{Spohn,Bleher,Tracy,Okounkov2,BH8,BH9}.

In this paper, in  view of these interesting applications, we study the  Airy
matrix model, and  higher Airy matrix models,  with a logarithmic potential,  in
detail.
We first obtain the Virasoro constraints for the
partition function $Z$ of (\ref{intro1}) ; this gives 
a series expansion 
 in terms of   the  parameters $t_n = {\rm tr} \Lambda^{-n-\frac{1}{2}}$, constructed
from the source matrix $\Lambda$. The parameters $t_n$ characterize the intersection numbers of the 
moduli space of curves and  the KdV hierarchies,  following Witten's well-known conjecture
\cite{Witten2,Kontsevich}.
The Virasoro constraints are derived from the equations of motion for this matrix model.
The remarkable difference with the Kontsevich model \cite{Gross} 
is that the equation of motion
becomes here a third order differential equation, due to the presence of the logarithmic
term. This differential equation leads to the appearance of a 
new series of parameters
$t_{n/2}$ ($n/2$ is a half-integer) in addition to the $t_{n}$ (n is an integer) of the 
KdV hierarchies. 
It is interesting to analyze the role of  these new  variables
$t_{n/2}$,  absent from the usual KdV hierarchies. 
These parameters $t_{n/2}$ (n/2 half-integer) correspond to the Ramond
sector in  string theory,  decoupled from the Neveu-Schwarz sectors
 in the Kontsevich model or in the theory of intersection numbers of Riemann surfaces. 
Parameters  similar to $t_{n/2}$ appear also  in the antisymmetric Gaussian random matrix theory
\cite{BH4aa}. 

In the large N limit the equation of motion leads to a  Riemann-Hilbert  integral equation. We have verified the consistency of the solution with the results derived from the Virasoro constraints.
The free energy is expressed through the parameters $t_{n}$ and $t_{\frac{n}{2}}$.
 Starting from the duality relation between the Gaussian two-matrix and the Kontsevich-Penner model,
we consider the Fourier transform of the correlation functions in the two-matrix model. The terms 
$t_{\frac{n}{2}}$ are generated by  the correlations
between
the edge of the spectrum and the bulk. The edge behavior and the bulk behavior obey different scaling laws.
This paper is organized as follows; in  section 2,  we recall how a  
logarithmic potential is generated from a duality relation for characteristic polynomials 
in the Gaussian two-matrix model. In  section 3, the equations of motion are derived
for the Kontsevich model (p=2) with a logarithmic potential. The Virasoro
constraints are obtained, as differential equations for $t_0$, $t_{\frac{1}{2}}$ and
$t_1$. From these differential equations, we construct the series expansion of the 
free energy $F$ in terms of the  $t_n$ and $t_{\frac{n}{2}}$. 
In  section 4, the integral equation (Riemann-Hilbert problem)
is solved for the Airy matrix model with a logarithmic potential. 
In  section 5, we discuss the replica method which provides explicit results for 
one marked point. The corresponding correlation function of the two-matrix model
is discussed and the result of its Fourier transform is compared with the result of the
free energy obtained from Virasoro constraints.
The section 6 is devoted to discussions. The  formula which expresses the $p$-th derivatives
with respect to the external source matrix $\Lambda$
in terms of its eigenvalues   is presented in the appendix A. In Appendix B,
 the equivalence between the unitary matrix model with an external source
and $(p=-2)$-higher Airy matrix model with a logarithmic potential is briefly sketched.
  
\vskip 5mm
\section{A logarithmic potential}
\setcounter{equation}{0}
\renewcommand{\theequation}{2.\arabic{equation}}

For the one-matrix model, we have shown that the Kontsevich model is related to a
 matrix model at the edge of its spectrum,  through a duality relation and the replica
method. 
For the mathematical definition of the intersection numbers of the moduli space of curves, we refer to \cite{Kontsevich} .  It involves an integration over the compactfied moduli space $\bar M_{g,n}$ with genus $g$ and n-marked
points,
\be\label{inter1}
< \tau_{d_1} \cdots \tau_{d_n} > = \int_{\bar M_{g,n}}
c_1({\mathcal{L}}_1)^{d_1} \cdots c_1({\mathcal{L}}_n)^{d_n}.
\ee
where $c_1$ is the first Chern class and ${\mathcal{L}}_i$ is a cotangent line bundle
at the i-th marked point.
 This definition  of the intersection numbers has been 
generalized \cite{Witten1} to the moduli space of $p$-spin curves. The intersection numbers 
have now an additional spin-index like in  $\tau_{n,j}$, in which $j$ takes values from $0$ to $p-1$.
They are defined by
\be
< \tau_{d_1,j_1} \cdots \tau_{d_n,j_n} > = \frac{1}{p^g}\int_{\bar M_{g,n}} 
c_D({\mathcal{V}})
c_1({\mathcal{L}}_1)^{d_1} \cdots c_1({\mathcal{L}}_n)^{d_n} .
\ee
where $c_D({\mathcal{V}})$ is a D-dimensional top Chern class of the vector 
bundle ${\mathcal{V}}$, ${\mathcal{V}}= H^{1}(\Sigma,{\mathcal{T}})$. 
There is a cover of Riemann surface, and the line bundle ${\mathcal{L}}$ has $p$ roots,
\be
{\mathcal{L}} \simeq {\mathcal{T}}^{\otimes p}
\ee
where ${\mathcal{T}}$ is isomorphism class. 
There are ($p-1$) roots corresponding to the Neveu-Schwarz sector 
and one  to the Ramond sector in  string theory.

These intersection numbers with spin indices
may be  computed from the $p$-th higher Airy matrix model.
Non-vanishing intersection numbers satisfy the  condition,
\be
 \sum_{i=1}^n d_i + D = 3 g - 3 + n
\ee

 In the limit $p\to-1$ the top Chern class becomes the Euler characteristics with marked points. In this case, the higher Airy matrix model
reduces to the Penner model (logarithmic potential), from which one computes the Euler characteristics 
$\chi_{g,n}$ \cite{HZ,Penner}. In this section, we will discuss the duality  
between the Kontsevich model  with a logarithmic potential (\ref{intro1}) and the  Gaussian 
two-matrix model with an external source.

In previous papers \cite{BH4,BH6}, we have discussed the time dependent ($c=1$)
Gaussian random matrix theory. We have shown that
the correlation function at two different times is equivalent to
the correlation function of a  two-matrix model with the
following probability distribution,
\be\label{PP}
P (M_1,M_2) = \frac{1}{Z}{\rm exp}\bigg( - \frac{1}{2}{\rm tr} M_1^2 - \frac{1}{2}{\rm tr} M_2^2
- c {\rm tr} M_1 M_2 + {\rm tr} M_1 A_1 + {\rm tr}M_2 A_2\bigg)
\ee
where $Z$ is a normalization constant. 

We consider the correlation function $F_{k_1,k_2}$
of  characteristic polynomials, which is defined by the $(k_1+k_2)$-point 
correlation function,
\be
F_{k_1,k_2} = < \prod_{\alpha=1}^{k_1}{\rm det}(\lambda_\alpha - M_1) \prod_{\beta=1}^{k_2}
{\rm det} (\mu_\beta - M_2) >
\ee
where the average $<\cdots>$ is computed with the distribution $P$ in (\ref{PP}).

Let us briefly review how this correlation function reduces to the Kontsevich 
model with a logarithmic potential \cite{BH4}. 
This $F_{k_1,k_2}$ is expressed as a Grassmann integral over  $\psi_\alpha (\alpha=1,...,k_1)$,
$\chi_\beta (\beta= 1,...,k_2)$,
\be
F_{k_1,k_2} = < \int d\bar \psi d \psi d\bar 
\chi d\chi e^{\bar \psi_\alpha (\lambda_\alpha - M_1)\psi_\alpha
+\bar\chi_\beta (\mu_\beta - M_2)\chi_\beta}>
\ee
The integration over the matrices $M_1$ and $M_2$ in then easy with the Gaussian distribution (\ref{PP}). 
It yields quartic terms in $\psi$ and $\chi$. These terms may be
expressed as integrals over Hermitian matrices $B_1$, $B_2$ and complex matrices $D$, $D^\dagger$.
($k_1\times k_1$ for the auxiliary matrix $B_1$, $k_2\times k_2$ for 
$B_2$, and $k_1\times k_2$ for the complex matrix $D$).
\ba
&&{\rm exp}[-\frac{N}{2(1-c^2)}\bar\psi \psi \bar \psi \psi] =
\int dB_1 {\rm exp}\bigg(-\frac{N}{2}{\rm tr}B_1^2 + \frac{iN}{\sqrt{1-c^2}}B_1 \bar \psi \psi
\bigg)\nonumber\\
&&{\rm exp}[-\frac{N}{2(1-c^2)}\bar\chi \chi \bar \chi \chi] =
\int dB_2 {\rm exp}\bigg(-\frac{N}{2}{\rm tr}B_2^2 + \frac{iN}{\sqrt{1-c^2}}B_2 \bar \chi 
\chi\bigg)\nonumber\\
&&{\rm exp}[\frac{Nc}{1-c^2}\bar\psi \chi \bar\chi \psi] = \int dDdD^\dagger
{\rm exp}\bigg(-N{\rm tr} D^\dagger D + \frac{N\sqrt{c}}{\sqrt{1-c^2}}
{\rm tr} (D\bar \psi \chi+ D^\dagger \bar\chi \psi)\bigg)\nonumber\\
\ea
Note that we have traded the integrations over $N\times N$ matrices by integrals over matrices whose sizes are given by $k_1$ and $k_2$. One can then  integrate out the Grassmann variables $\psi$ and $\chi$, and
$F_{k_1,k_2}$ is expressed as
\be
F_{k_1,k_2} = \int dB_1 dB_2 d D d D^{\dagger}
e^{-\frac{N}{2} {\rm tr} (B_1^2+B_2^2 + 2 D^\dagger D)+ N {\rm tr} {\rm log}( 1 - X)}
\ee
where
\be\label{Xi}
X = \left( \matrix{ 
\frac{i\sqrt{1-c^2}}{a_1- c a_2} \tilde B_1& 
\frac{\sqrt{c(1-c^2)} }{a_1- c a_2} D\cr
 \frac{\sqrt{c(1-c^2)}}{a_2-c a_1} D^{\dagger}
& \frac{i\sqrt{1-c^2}}{a_2 - c a_1} 
\tilde B_2
\cr } \right).
\ee
with
\ba
 &&(\tilde B_1)_{\alpha\alpha'} = (B_1)_{\alpha\alpha'} - i\sqrt{1-c^2}\lambda_{\alpha} \delta_{\alpha\alpha'}\nonumber\\
&&(\tilde B_2)_{\beta\beta'} =( B_2)_{\beta\beta'} - i\sqrt{1-c^2}\mu_{\beta} \delta_{\beta\beta'}\nonumber\\
\ea
We have assumed that the external source matrices $A_1$ and $A_2$ are multiple of the identity $A_1= a_1\cdot I,A_2=a_2\cdot
I$.
Introducing the diagonal matrices $\Lambda_1$ and $\Lambda_2$ 
\be
\Lambda_1= {\rm diag}(\lambda_1,...,\lambda_{k_1}),
\hskip 5mm
\Lambda_2 = {\rm diag}(\mu_1,...,\mu_{k_2})
\ee
we obtain
\ba
F_{k_1,k_2}&=& e^{\frac{N}{2}(1-c^2) {\rm tr} (\Lambda_1^2+\Lambda_2^2)}\int
d\tilde B_1 d\tilde B_2 dD dD^\dagger
e^{-\frac{N}{2}{\rm tr} (\tilde B_1^2+\tilde B_2^2+ 2 D^\dagger D)+ N {\rm tr}{\rm log}(1-X)
}\nonumber\\
&\times& e^{- i N \sqrt{1-c^2} {\rm tr} \tilde B_1 \Lambda_1- i N \sqrt{1 - c^2} {\rm tr}
\tilde B_2 \Lambda_2}
\ea
We now restrict ourselves to  $a_2 = 0$. From the expression of $X$, we have
\be
 {\rm tr} X^2 = - \frac{(1-c^2)}{a_1^2} {\rm tr} \tilde B_1^2 - \frac{1-c^2}{c^2 a_1^2}{\rm tr} \tilde B_2^2
- \frac{2(1-c^2)}{a_1^2}{\rm tr} D^\dagger D
\ee
If we take $a_1^2 = 1-c^2$, the quadratic term in $\tilde B_1$ is cancelled.
The quadratic term in $\tilde B_2$ does not vanish, it becomes
$-\frac{N}{2}(1-c^{-2}){\rm tr} \tilde B_2^2$.
The quadratic term ${\rm tr}D^\dagger D$ is also cancelled.

We now denote $\tilde B_1,\tilde B_2$ by $B_1,B_2$.
We find
\ba
{\rm tr} X^3 &=& - \frac{i(1-c^2)^{3/2}}{a_1^3}{\rm tr}B_1^3
- \frac{3 i (1-c^2)^{3/2}}{a_1^3}{\rm tr} D D^\dagger B_1
+ \frac{3 i (1-c^2)^{3/2}}{c a_1^3}{\rm tr} D^\dagger  D B_2\nonumber\\
&+& \frac{i (1-c^2)^{3/2}}{c^3 a_1^3}{\rm tr} B_2^3
\ea

Given the  factor $N$
in the exponent, the edge scaling limit under consideration corresponds  to \be\label{order}
B_1\sim O(N^{-\frac{1}{3}}), B_2\sim O(N^{-\frac{1}{2}}), D\sim O(N^{-\frac{1}{3}})
\ee
in the large N limit, since the quadratic term in $B_2$ does not vanish.
In this limit most terms disappear ; for instance
\be
N {\tr} (D^\dagger D B_2)\sim N^{-\frac{1}{6}}
\ee
is negligible.
Then, in the large N limit (\ref{order}), 
 after dropping the negligible terms, we obtain
\be
F_{k_1,k_2} = \int dB_1 dB_2 d D^\dagger d D e^{- i N {\tr} B_1 \Lambda_1 - i N {\tr} B_2 \Lambda_2
+ \frac{i}{3}N {\tr} B_1^3 - \frac{N}{2}(1- \frac{1}{c^2}) {\tr} B_2^2 + i N {\tr}
(D D^\dagger B_1)}
\ee
Since the matrix $B_2$  is decoupled, we can  integrate it out. 
Then, dropping the contribution from the integral over $B_2$, 
we find   
\be
F_{k_1,k_2} = \int dB_1 dD^\dagger dD
 e^{-i {\tr}B_1 \Lambda_1 + \frac{i}{3}{\tr} B_1^3 + i {\tr} D D^\dagger B_1}
\ee
where we have absorbed the powers of  $N$ in a rescaling.

We may now  integrate out the matrices $D$ and $D^\dagger$ ($D$ is a $k_1\times k_2$ complex
matrix); this yields a one matrix integral with a  logarithmic potential,
\be\label{log}
F_{k_1,k_2} = \int dB_1 e^{\frac{i}{3}{\tr} B_1^3 - k_2 {\tr}{\rm log} B_1 - i {\tr} B_1 \Lambda_1}.
\ee
where $B_1$ is a  $k_1\times k_1$ Hermitian matrix. If we chose the replacement $B_1\to - i B$, we obtain 
the model (\ref{intro1}).

\vskip 3mm

\section{ Virasoro constraints}
\setcounter{equation}{0}
\renewcommand{\theequation}{3.\arabic{equation}}
The Kontsevich model with a logarithmic potential,
\be
Z = \int dB e^{{\rm tr}(- \frac{1}{3} B^3 + \Lambda B + k {\rm log} B)},
\ee
in which $B$ is an Hermitian $P\times P$ matrix (we have replaced $k_1$ by $P$ and $k_2$ by $k$). 
satisfies the trivial equations of motion,
\be\label{eqM}
\int dB \frac{\partial}{\partial B_{ba}} 
e^{{\rm tr}(- \frac{1}{3} B^3 + \Lambda B + k {\rm log} B)} =0 \ee
from which one obtains readily 
\be
\bigg(-\bigg(\frac{\partial}{\partial \Lambda}\bigg)_{ab}^3 + \bigg(\Lambda^T 
\frac{\partial}{\partial \Lambda}\bigg)_{ab} + (P + k) \delta_{ab} \bigg) Z = 0.
\ee

Since $Z$ is a function of the  eigenvalues $\lambda_i$ of $\Lambda$, one can trade this for differential equations in terms of these eigenvalues (see appendix A), 
\ba\label{virasoro}
&&\frac{\partial^3 Z}{\partial {\lambda_c}^3} + \sum_{d\ne c} \frac{1}{\lambda_c-\lambda_d}
(\frac{\partial}{\partial \lambda_c} - \frac{\partial}{\partial \lambda_d})
(2 \frac{\partial}{\partial \lambda_c} + \frac{\partial}{\partial \lambda_d}) Z
\nonumber\\
&&- \sum_{d\ne c}\frac{1}{(\lambda_c- \lambda_d)^2}
(\frac{\partial}{\partial \lambda_c}-\frac{
\partial}{\partial \lambda_d})Z
+ 2 \sum_{d\ne e,c} \frac{1}{(\lambda_c -\lambda_e)(\lambda_e - \lambda_d)}(\frac{\partial}{
\partial \lambda_c}-\frac{\partial}{\partial \lambda_e})\nonumber\\
&&-\lambda_c \frac{\partial Z}{\partial \lambda_c} - (P + k ) Z = 0
\ea

The zero-th order contribution for large $\Lambda$,  is obtained from the shift  $B \to B + \Lambda^{\frac{1}{2}}$ ; then keeping only the terms which grow for large $\Lambda$ one finds 
\ba\label{Z0}
 Z_0 &=& \int d B e^{- {\rm tr} B^2 \Lambda^{\frac{1}{2}}
+ \frac{2}{3}{\rm tr} \Lambda^{\frac{3}{2}} + \frac{k}{2}{\rm tr}{\rm log} \Lambda}\nonumber\\
&=& \frac{1}{\prod_{i,j} ( \sqrt{\lambda_i} + \sqrt{\lambda_j} )^{\frac{1}{2}}}
e^{\frac{2}{3}\sum \lambda_i^{\frac{3}{2}}}\prod_i \lambda_i^{\frac{k}{2}}\ea
In the limit  $\lambda_i \to \infty$, the partition function reduces to $Z_0$.
Therefore, the partition function $Z$  may be expressed as
\be\label{g}
Z = Z_0 g(\lambda)
\ee
where $g$ has an expansion in  inverse powers of $\sqrt{\lambda}$ :
\be
g = 1 + O(\frac{1}{\lambda^{\frac{3}{2}}})
\ee

The Virasoro constraints (\ref{virasoro}) lead to a sequence of equations, which fix  the   coefficients of the terms $\lambda_c^{-\frac{n}{2}}$.
Let us thus write the Virasoro constraints in terms of the function $g$ of (\ref{g}). For this purpose, we have to substitute $Z_0$
into (\ref{virasoro}). The resulting equations are cumbersome. To avoid complicated and long
expressions, we take 
 the simple case of $P=2$. Although this simple case is manifestly not sufficient
to  determine the expansion in terms of the $t_n$, it is 
instructive and useful also for arbitrary $P$ as shown below. Then for $P=2$, the equations 
(\ref{virasoro}) become 
\ba
&&\bigg(\frac{\partial^3}{\partial \lambda_1^3} + \frac{1}{\lambda_1-\lambda_2}(\frac{\partial}{
\partial \lambda_1}-\frac{\partial}{\partial \lambda_2})(2 \frac{\partial}{\partial \lambda_1}+
\frac{\partial}{\partial \lambda_2}) \nonumber\\
&&- \frac{1}{(\lambda_1-\lambda_2)^2}
(\frac{\partial}{\partial \lambda_1}-\frac{\partial}{\partial \lambda_2}) - \lambda_1 
\frac{\partial}{\partial \lambda_1}- (2 + k )\bigg) Z = 0
\ea
Using $Z = Z_0 g$, we obtain the equations for $g$,
\be\label{virasoro2}
a_1 g + a_2 \frac{\partial g}{\partial \lambda_2} + a_3 \frac{\partial g}{\partial \lambda_1}
+ a_4 \frac{\partial^2 g}{\partial \lambda_1^2} 
 + a_5 \frac{\partial^2 g}{\partial \lambda_1 \partial \lambda_2}
+ a_6 \frac{\partial^2 g}{\partial \lambda_2^2} + a_7 \frac{\partial^3}{\partial \lambda_1^3}
= 0
\ee
where
\ba
&&a_1 = \frac{1}{\sqrt{\lambda_1} \lambda_2}\bigg(\frac{1 - 2 k}{4}\bigg) 
+ \frac{1}{\lambda_1}
\bigg( \frac{1 - 2 k }{2 \sqrt{\lambda_2}}- \frac{(2 k -1)(2 k -5)}{16 \lambda_2^2} \bigg)
\nonumber\\
&&+ \frac{1}{\lambda_1^{\frac{3}{2}}} \bigg(\frac{5 - 24 k + 12 k^2}{16} +
 \frac{k-1}{2 {\lambda_2}^{\frac{3}{2}}} \bigg)
+ \frac{1}{{\lambda_1}^2}\bigg( - \frac{(k-1)(2 k-3)}{4 \lambda_2} \bigg)\nonumber\\
&&+ \frac{1}{{\lambda_1}^{\frac{5}{2}}}\bigg( \frac{k-1}{{\lambda_2}^{\frac{1}{2}}} \bigg)
+ \frac{1}{{\lambda_1}^3} \bigg( \frac{(2 k-1)(2 k- 5)(2 k- 9)}{64} \bigg)
\nonumber\\
&&+ \frac{-1 + 16 k - 12 k^2}{32 {\lambda_1}^{\frac{5}{2}}} \frac{1}{\sqrt{\lambda_1}+
\sqrt{\lambda_2}}
\ea
\ba\label{a2}
&&a_2 =  \frac{1}{2} \frac{1}{\sqrt{\lambda_1}+\sqrt{\lambda_2}} - \frac{3}{2}\cdot
\frac{1}{\sqrt{\lambda_1}-\sqrt{\lambda_2}} + \frac{1 - 2 k}{2 \lambda_1 \lambda_2}
+ \frac{1}{{\lambda_1}^{\frac{3}{2}} \sqrt{\lambda_2}} \nonumber\\
&&- \frac{3 k }{4 {\lambda_1}^{\frac{3}{2}}} \frac{1}{\sqrt{\lambda_1}+\sqrt{\lambda_2}}
+ \frac{1 - \frac{3}{4} k}{{\lambda_1}^{\frac{3}{2}}} 
\frac{1}{\sqrt{\lambda_1}-\sqrt{\lambda_2}}
+ \frac{1}{4 \lambda_1 (\sqrt{\lambda_1} - \sqrt{\lambda_2})^2}
\ea
\ba\label{a3}
&&a_3 = 2 \lambda_1 + \frac{3 k}{\sqrt{\lambda_1}} 
-  \frac{1}{2} \frac{1}{\sqrt{\lambda_1}+\sqrt{\lambda_2}} 
+ \frac{3}{2}\cdot
\frac{1}{\sqrt{\lambda_1}-\sqrt{\lambda_2}}\nonumber\\
&&+ \frac{15-36 k + 12 k^2}{16 {\lambda_1}^2} + \frac{1 - 2 k}{4 \lambda_1 \lambda_2}
\nonumber\\
&&-\frac{3 k }{4 {\lambda_1}^{\frac{3}{2}}} \frac{1}{\sqrt{\lambda_1}+\sqrt{\lambda_2}}
- \frac{1 - \frac{3}{4} k}{{\lambda_1}^{\frac{3}{2}}} 
\frac{1}{\sqrt{\lambda_1}-\sqrt{\lambda_2}}
- \frac{1}{4 \lambda_1 (\sqrt{\lambda_1} - \sqrt{\lambda_2})^2}
\ea
\be\label{a4}
a_4 =  3 \sqrt{\lambda_1} - \frac{3 (1 - 2 k)}{4 \lambda_1}
- \frac{1}{2 \sqrt{\lambda_1} (\sqrt{\lambda_1}+\sqrt{\lambda_2})}
+\frac{1}{\sqrt{\lambda_1}(\sqrt{\lambda_1}-\sqrt{\lambda_2})}
\ee
\be
a_5=a_6= - \frac{1}{\lambda_1-\lambda_2}
\ee
\be
a_7 = 1
\ee

We now return to general $P$ (not simply $P=2$) and define  the parameters $t_n$ as
\be\label{tn}
t_n = \sum_{i=1}^P \frac{1}{{\lambda_i}^{n + \frac{1}{2}}}
\ee
in which $n$ takes both integer and half-integer values  ( $n=0,\frac{1}{2},1, \frac{3}{2}, \cdots$ ).
Note that only integers appear in the Kontsevich model. The appearance of half-integers
is a characteristic of  the present model with a logarithmic potential.
The derivatives with respect to  $\lambda_j$ are replaced by
\be
\frac{\partial}{\partial \lambda_j} = \sum_n \frac{\partial t_n}{\partial \lambda_j}
\frac{\partial}{\partial t_n} = 
 - \sum_n (n + \frac{1}{2}) \frac{1}{{\lambda_j}^{n + \frac{3}{2}}}
\frac{\partial}{\partial t_n}
\ee
\be
\frac{\partial^2}{\partial {\lambda_j}^2}= 
\sum_n \frac{(n+\frac{1}{2})(n+ \frac{3}{2})}{
{\lambda_j}^{n + \frac{5}{2}}}
 \frac{\partial}{\partial t_n}+ \sum_n \sum_m \frac{(n+\frac{1}{2})
(m+ \frac{1}{2})}{{\lambda_j}^{m+n + 3}}
\frac{\partial^2}{\partial t_n \partial t_m}
\ee
\be
\frac{\partial^2}{\partial \lambda_1 \partial \lambda_2} = \sum_n \sum_m 
\frac{(n + \frac{1}{2})(m+ \frac{1}{2})}{{\lambda_1}^{n+\frac{3}{2}} {\lambda_2}^{m+\frac{3}{2}}}
\ee
\ba
\frac{\partial^3}{\partial {\lambda_1}^3} &=&
- \sum_n \frac{(n+\frac{1}{2})(n+ \frac{3}{2})(n + \frac{5}{2})}{{\lambda_1}^{n + \frac{7}{2}}}
\frac{\partial}{\partial t_n}\nonumber\\
&-&\sum_n \sum_m \frac{(n+ \frac{1}{2})(m+ \frac{1}{2})( 2 n + m + \frac{9}{2})}{
{\lambda_1}^{n+m+4}}\frac{\partial^2}{\partial t_n \partial t_m}\nonumber\\
&-& \sum_n \sum_m \sum_j \frac{(n+\frac{1}{2})(m+\frac{1}{2})
(j+\frac{1}{2})}{{\lambda_1}^{n+m+j+\frac{9}{2}}}\frac{\partial^3}{\partial t_n \partial t_m
\partial t_j}
\ea
where $n,m,j = 0,\frac{1}{2},1,\frac{3}{2},2,\cdots$.
\vskip 5mm

Returning now to $P=2$, at lowest order in the $1/\sqrt{\lambda_1}$ expansion, $a_1$ becomes
\be
a_1 \sim \frac{1}{\sqrt{\lambda_1}}\bigg( \frac{1}{4}(\frac{1}{\sqrt{\lambda_1}}+
\frac{1}{\sqrt{\lambda_2}})^2 - \frac{k}{2}(\frac{1}{\lambda_1}+\frac{1}{\lambda_2})\bigg)
+O(\frac{1}{\lambda_1})
\ee
This may be expressed in terms of the $t_n$ in (\ref{tn}) as,
\be
a_1 \sim \frac{1}{\sqrt{\lambda_1}}(\frac{1}{4}t_0^2 - \frac{k}{2}t_1)
\ee

\be
a_2 \frac{\partial}{\partial \lambda_2} \sim - \frac{1}{\sqrt{\lambda_1}}
( \frac{1}{2}t_1\frac{\partial}{\partial t_0} + t_{\frac{3}{2}}\frac{\partial}{\partial t_{\frac{1}{2}}}+ \cdots)
\ee

\be
a_3 \frac{\partial}{\partial \lambda_1} \sim 2 \lambda_1 \frac{\partial}{\partial \lambda_1}\sim
-\frac{1}{\sqrt{\lambda_1}}\frac{\partial}{\partial t_0}
\ee
We have considered only the case $P=2$, but this simple calculation is enough to determine 
correctly the
coefficients of $t_0^2$ and $t_1$, 
which are defined as the sum of $\lambda_i^n$ up to  
$i=P$ as (\ref{tn}).

The coefficients $a_4,a_5,a_6,a_7$ do not appear yet at this order since the multiplications of derivatives in (\ref{virasoro2})
give higher orders in $\lambda_1^{-1}$. Then, we obtain the first equation of order $\lambda_1^{-1/2}$,
\be
\bigg(-\frac{\partial}{\partial t_0} + \frac{1}{4}t_0^2 -\frac{k}{2}t_{\frac{1}{2}} + \sum_{n=0,\frac{1}{2},
1,...} (n + \frac{1}{2}) t_{n+1}\frac{\partial}{\partial t_n}\bigg)g = 0
\ee

Using $F = {\rm log} g$, it becomes
\be
\frac{\partial F}{\partial t_0} = \frac{1}{4}t_0^2 - \frac{k}{2} t_{\frac{1}{2}} + 
\sum_{n=0,\frac{1}{2},1,..} (n+ \frac{1}{2}) t_{n+1}\frac{\partial F}{\partial t_n}
\ee

For the next order $\lambda_1^{-1}$, we need to evaluate   $a_1$ for $N=3$, since the 
$N=2$ results are not sufficient to determine the coefficients of $t_n$ which appear 
in the equation.
We obtain
\ba
a_1&=& \frac{1}{\sqrt{\lambda_1}}\bigg( (\frac{1}{4}- \frac{k}{2}) (\frac{1}{\lambda_2}+
\frac{1}{\lambda_3}) + \frac{1}{2\sqrt{\lambda_2 \lambda_3}}\bigg)\nonumber\\
&+& \frac{1}{\lambda_1}\bigg((\frac{1}{2}-k)(\frac{1}{\sqrt{\lambda_2}}+\frac{1}{\sqrt{\lambda_3}})
+\frac{(-5 + 12 k - 4 k^2)}{16}(\frac{1}{\lambda_2^2}+ \frac{1}{\lambda_3^2})\nonumber\\
&+& (-\frac{1}{2}+ \frac{k}{2})(\frac{1}{\lambda_2 \lambda_3}+ \frac{1}{\lambda_3^{\frac{3}{2}}\sqrt{\lambda_2}}+
\frac{1}{\lambda_2^{\frac{3}{2}}\sqrt{\lambda_3}})\bigg)
\ea
From (\ref{a2}),(\ref{a3}) and (\ref{a4}), we obtain the second equation,  involving now a  derivative with respect to the
 $t_{\frac{1}{2}}$,

\ba\label{L(-1/2)}
&&\bigg( -2 \frac{\partial}{\partial t_{\frac{1}{2}}} - k t_0 
+ \frac{k}{4} {t_{\frac{1}{2}}}^2
+ \frac{k}{2} t_0 t_1 - \frac{1}{4} {t_0}^2 t_{\frac{1}{2}} 
- \frac{1}{16} t_{\frac{3}{2}}
-\frac{1}{4} k^2 t_{\frac{3}{2}}\nonumber\\
&&- \sum_{n=0,\frac{1}{2},1,..}  (2 n + 1) t_{n + \frac{1}{2}} \frac{\partial}{\partial t_n}
 + k \sum_{n=0,\frac{1}{2},1,..} ( n + \frac{1}{2}) t_{n+2} \frac{\partial}{\partial t_n}
\nonumber\\
&&- \frac{1}{2} \sum_{-i -j + k = -\frac{3}{2}} (k + \frac{1}{2})
t_i t_j \frac{\partial}{\partial t_k}
- \frac{1}{2} \sum_{-i + j + k =-\frac{5}{2}} ( j+\frac{1}{2})(k + \frac{1}{2}) t_i 
\frac{\partial^2}{\partial t_j \partial t_k}\bigg) g = 0\nonumber\\
\ea

The next order  is proportional to $ {\lambda_1}^{-\frac{3}{2}} $,  and we obtain
\ba\label{L(0)}
&&\bigg( - 3 \frac{\partial}{\partial t_1} - \frac{1}{16}  - \frac{3}{4} k^2 + 
\frac{k-1}{2} t_0 t_{\frac{1}{2}} \nonumber\\
&& -  \sum_{n=0,\frac{1}{2},1,..} (\frac{1}{2} + n) t_n \frac{\partial}{\partial t_n}
- \sum_{n=0,\frac{1}{2},1,..} (n + \frac{1}{2}) t_{n + \frac{3}{2}} \frac{\partial}{\partial t_n} \bigg) 
g = 0                                
\ea

These equations determine the free energy  $F= {\rm log} g$ as
 up to order $O(\lambda^{-\frac{9}{2}})$,
\ba\label{Ft}
  F &=& \frac{1}{12} t_0^3 + \frac{1}{48} t_1 + \frac{1}{2} k t_0 t_{\frac{1}{2}} 
+ \frac{1}{4} k^2 t_1
\nonumber\\
&+& \frac{1}{24} t_0^3 t_1 + (\frac{1}{192} + \frac{1}{16} k^2) t_1^2
+ \frac{1}{4} k t_0 t_{\frac{1}{2}} t_1 + \frac{1}{24} k t_{\frac{1}{2}}^3 + (\frac{1}{32}+
\frac{3}{8} k^2)t_0 t_2 + \frac{1}{4} k t_0^2 t_{\frac{3}{2}} \nonumber\\
&+& \frac{1}{4} k^2 t_{\frac{1}{2}}t_{\frac{3}{2}} + \frac{1}{6}(k+k^3) t_{\frac{5}{2}} \nonumber\\
&+&  \frac{1}{64} t_0^4 t_2 + \frac{1}{6} k t_0^3 t_{\frac{5}{2}} + \frac{1}{48} t_0^3 t_1^2
+ (\frac{5}{128} + \frac{15}{32} k^2) t_0^2 t_3\nonumber\\
&+& \frac{3}{16} k t_0^2 t_{\frac{1}{2}} t_2 + \frac{1}{4} k t_0^2 t_1 t_{\frac{3}{2}}
+ \frac{1}{2}( k +  k^3) t_0 t_{\frac{7}{2}}
+ \frac{1}{12} k^2 t_0 t_{\frac{1}{2}} t_{\frac{5}{2}}\nonumber\\
&+& (\frac{1}{32} + \frac{3}{8} k^2) t_0 t_1 t_2 
+ \frac{1}{4} k^2 t_0 t_{\frac{3}{2}}^2 + \frac{1}{8} k t_0 t_{\frac{1}{2}}^2 t_{\frac{3}{2}}
+ \frac{1}{8} k t_0 t_{\frac{1}{2}} t_1^2\nonumber\\
&+& \frac{1}{14}(\frac{73}{32} k + \frac{43}{8} k^3) t_{\frac{1}{2}}t_3
+ \frac{3}{16} k^2 t_{\frac{1}{2}}^2 t_2 + \frac{1}{4} k^2 t_{\frac{1}{2}} t_1 t_{\frac{3}{2}}
+ \frac{1}{6}( k +  k^3) t_1 t_{\frac{5}{2}}\nonumber\\
&+& (\frac{1}{576} + \frac{1}{48} k^2) t_1^3 + (\frac{1}{8} k + \frac{1}{4} k^3) t_{\frac{3}{2}}
t_2 + \frac{1}{9}(\frac{105}{1024} + \frac{607}{128} k^2 + \frac{169}{64} k^4) t_4\nonumber\\
&+& \frac{1}{24} k t_{\frac{1}{2}}^3 t_1
\ea

This expression is consistent with the previous result \cite{BH4}. Note that
the parameters $t_n$ with the half-integers $n$, $t_{\frac{1}{2}},t_{\frac{3}{2}},..$,
appear together with the coefficients proportional to $k$. When $k$ goes to zero, the free
energy $F$ of (\ref{Ft}) reduces to the Kontsevich free energy.
Another remarkable propertiy of (\ref{Ft}) is that when $k$ is of order  $P$, 
many terms are of the same order in the large P limit. The leading order is $P^2$ which gives
genus zero contributions. We will discuss the large P limit in a  later 
section from a different approach based on integral equations.

To express these equations in  compact form, it is  convenient to introduce
the differential operators $J_n^{(k)}$, obtained as follows \cite{Adler}.
 \be
{J_m}^{(1)}(x) = \frac{\partial}{\partial x_m} - m x_{-m},\hskip 5mm 
(m= ...,-2,-1,0,1,2,...)
\ee
and $x_m=0$ for $x\ge 0$.
We define $J_m^{(k)}$ ($k>1$) from $J_m^{(1)}$ as
\be
J_m^{(2)} = \sum_{i+j=m}: J_i^{(1)} J_j^{(1)}:
\ee
where $: \cdots :$ means  normal ordering, i.e. 
pulling the differential operator
 to the right.
Then we obtain
\be
{J_m}^{(2)} = \sum_{i+j=m}\frac{\partial^2}{\partial x_i \partial x_j} + 
2 \sum_{-i +j = m} i x_i \frac{\partial}{\partial x_j} + 
\sum_{-i-j=m} (i x_i)(j x_j)
\ee
\ba\label{J3}
{J_m}^{(3)} &=& \sum_{i+j+k=m} :J_i^{(1)} J_j^{(1)} J_k^{(1)} :
\nonumber\\
&=&\sum_{i+j+k=m}\frac{\partial^3}{\partial x_i \partial x_j \partial x_k}
+ 3 \sum_{-i+j+k = m} i x_i \frac{\partial^2}{\partial x_j \partial x_k}\nonumber\\
&& + 3 \sum_{-i-j+k=m}(i x_i)(j x_j) \frac{\partial}{\partial x_k} + \sum_{-i-j-k=m}
(i x_i)(j x_j)(k x_k)
\ea
where  $i,j,k= 1,2,3,...$.

By setting
\be
x_n = \frac{1}{n}t_{\frac{n-1}{2}},
\ee
we find
\be
J_{-4}^{(2)} = 2 t_0 t_1 + t_{\frac{1}{2}}^2 + 4 \sum_{n=0,\frac{1}{2},1,..} (n + \frac{1}{2})
t_{n+2} \frac{\partial}{\partial t_n}
\ee
\be
J_{-2}^{(2)} = t_0^2 + 2 \sum_{n=0,\frac{1}{2},1,..} (2 n + 1) t_{n+1}\frac{\partial}{\partial t_n}
\ee
\be
J_{-1}^{(2)} = 4 \sum_{n=0,\frac{1}{2},1,..}(n + \frac{1}{2}) t_{n+ \frac{1}{2}}
 \frac{\partial}{\partial t_n}
\ee
\be
J_0^{(2)} = 4 \sum_{n=0,\frac{1}{2},1,..}(n + \frac{1}{2})
 t_n \frac{\partial}{\partial t_n}
\ee
From (\ref{J3}), we have
\ba
J_4^{(3)} &=& 3 t_0^2 t_{\frac{1}{2}} + 3 \sum_{-i+j+k=-\frac{5}{2}}(2j+1)(2k+1) t_i
\frac{\partial^2}{\partial t_j \partial t_k}\nonumber\\
&+& 3 \sum_{-i-j+k = -\frac{3}{2}}(k + \frac{1}{2}) t_i t_j \frac{\partial}{\partial t_k}
\ea

Then, the first equation for the Virasoro constraints is expressed by
\be
\bigg(-\frac{\partial}{\partial t_0} + 
\frac{1}{4}J_{-2}^{(2)} - \frac{k}{2} t_{\frac{1}{2}}
\bigg) g = 0
\ee

The second  equation becomes
\be
\bigg( - 2\frac{\partial}{\partial t_{\frac{1}{2}}} - k t_0 -\frac{1}{16}t_{\frac{3}{2}}
-\frac{k^2}{4}t_{\frac{3}{2}} - \frac{1}{12}J_{-4}^{(3)} + \frac{k}{4}J_{-4}^{(2)} - \frac{1}{2}
J_{-1}^{(2)}
\bigg) g = 0
\ee

The third equation is expressed by
\be
\bigg( - 3\frac{\partial}{\partial t_1} - \frac{1}{16} 
- \frac{3}{4} k^2 + k t_0 t_{\frac{1}{2}}
- \frac{1}{4} {J_0}^{(2)} - \frac{1}{4} {J_{-3}}^{(2)} \bigg) g = 0
\ee

The differential operator $J_m^{(3)}$ appears only for the equation of order
$\lambda_1^{-n}$ (n=1,2,3,...). This is similar to the p-spin generalized Kontsevich model
without logarithmic term, where spin 0 equations are described by $J_n^{(2)}$ 
and the spin non-zero equation of motion is described by $J_m^{(3)}$ \cite{Dijkgraf}.

If we denote the differential operator $\frac{1}{4}{J_{2m}}^{(2)}$  as $L_m$ :\be
L_n = \frac{1}{4} J_{2n}^{(2)}
\ee
those $L_n$ have the commutation relations
\be
[ L_n, L_m ] = (n - m) L_{n+m}
\ee

\vskip 5mm

\section{Integral equation for the Airy matrix model }
\setcounter{equation}{0}
\renewcommand{\theequation}{4.\arabic{equation}}

For the unitary matrix model  the 
large N limit may be  solved by a Riemann-Hilbert integral equation \cite{BG}. We
apply here the same technique to the Kontsevich model with a logarithmic potential. 

When k=0 (the Kontsevich model), the equation of motion reduces to a simpler second order equation. 
Let us  first consider the
$k=0$ case as an exercise, following \cite{BG}.
\be\label{Konts}
\frac{\partial^2 Z}{\partial \lambda_c^2} 
+ \sum_d \frac{1}{\lambda_c - \lambda_d} \bigg(
\frac{\partial}{\partial \lambda_c} - \frac{\partial}{\partial \lambda_d} \bigg) Z
- \lambda_c  Z = 0
\ee
Changing to the free energy $W$
\be
Z = e^{P W}
\ee
(the original Gaussian matrices were $N\times N$, but the dual matrices are $P\times P$), 
and 
\be
\frac{\partial Z}{\partial \lambda_c} = 
P \bigg(\frac{\partial W}{\partial \lambda_c}\bigg)
Z = P W_c Z.
\ee
Introducing the density of eigenvalues
\be \rho(x) =\frac {1}{P} \sum_a \delta(x-\la_a) \ee
we consider $W$ as a functional of $\rho$  from which one obtains $W_a$ as 
\be W a = w(x)\vert_{x=\la_a} \ee
with 
\be w(x)= \frac{1}{P} \frac{d}{dx} \frac {\delta W}{\delta \rho(x)} \ee
 The second derivative in (\ref{Konts}) leads to two terms, but in the large P-limit the leading one is simply $w(x)^2$, leading to the integral equation 
\be\label{integraleq1}
w^2(x) + \int_a^b dy \rho(y) \frac{w(x)-w(y)}{x-y} = x
\ee
We define $f$ and $F$ as
\be
f(z) = \int_a^b dx \frac{\rho(x)}{z - x}
\ee
\be
F(z) = \int_a^b dx \frac{\rho(x)w(x)}{z-x}.
\ee
Inside the cut $z\in [a,b]$, 
\be
{\rm Re} F(z) = w^2(z) + w(x){\rm Re} f(z) - z
\ee
and for $z\in[-\infty,\infty]$,
\be
{\rm Im} F(z) = w(z) {\rm Im} f(z)
\ee
We make the ansatz 
\be\label{ansatz1}
F(z) = w^2(z) + w(z) f(z) - z
\ee
leading to 
\be\label{Im1}
{\rm Im} w ({\rm Re} f + 2 {\rm Re} w) = 0 \hskip 5mm z\in[-\infty,\infty], 
\ee
\be\label{Im2}
({\rm Im} w )({\rm Im }w + {\rm Im} f) = 0 \hskip 5mm z\in[a,b]
\ee
From (\ref{ansatz1}), in the $z\to \infty$ limit, we find
$F\sim 1/z$, $f(z)\sim 1/z$, and
\be\label{asympto}
w(z) = \sqrt{z} - \frac{1}{2 z} + O(z^{-3/2})
\ee
Since ${\rm Im} w \ne 0$ for $z\in [-\infty,-c]$, we have from (\ref{Im1}),
\be
{\rm Re} w = - \frac{1}{2} {\rm Re} f
\ee
This is equivalent to
\be
{\rm Im} (w(z) \sqrt{z + c}) = -\frac{1}{2}f(z) \sqrt{-z-c} \hskip 5mm (z\in [-\infty,-c])
\ee
Then by dispersion relation, we get
\ba
w(z) \sqrt{z + c} &=& -\frac{1}{2} \int dy \frac{f(y) \sqrt{-y-c}}{z-y}\nonumber\\
 &=& -\frac{1}{2\pi}\int_{-\infty}^{-c} dy \int_a^b dx \frac{\rho(x) \sqrt{-y-c}}{(y-x)(z-y)}
\ea
Noting that
\ba
&&\int_{-\infty}^{-c} dy\frac{\sqrt{-y - c}}{(y-x)(z-y)}
= \int_c^{\infty} dt \frac{\sqrt{t - c}}{(t+x)(t+z)}\nonumber\\
&=&  \frac{\pi}{\sqrt{z+c}+ \sqrt{x + c}}
\ea
Adding the integral constant $z + \frac{c}{2}$, which is determined from the asymptotic behavior
of (\ref{asympto}), we get
\be
w(z)\sqrt{z+c} = z + \frac{c}{2} - 
\frac{1}{2}\int_a^b dy \frac{\rho(y)}{\sqrt{z+c}+\sqrt{y+c}}
\ee
and the parameter $c$ is determined from the condition that there is no pole at $z=-c$,
\be\label{cs}
c = - \int_a^b dy \frac{\rho(y)}{\sqrt{y + c}}
\ee
Thus we obtain 
\be\label{wsol}
w(z) = \sqrt{z + c} + \frac{1}{2} \int_a^b dy \frac{\rho(y)}{(\sqrt{z+c} + \sqrt{y+c})\sqrt{
y + c}}
\ee

This  function $w(x)$ is indeed a solution of the integral equation (\ref{integraleq1}). 
The square of the second term, the part of the integration in (\ref{wsol}),
cancels with the second term of (\ref{integraleq1}), and the parameter $c$ is given
by (\ref{cs}).

By further  integration over $\rho(x)$,
we find the free energy $W$,
\ba\label{W}
W &=& \frac{2}{3}\int_a^b dz \rho(z) (z + c)^{\frac{3}{2}} - c \int_a^b dz \rho(z) \sqrt{z+c}\nonumber\\
&-& \int_a^b dz \int_a^b dy \rho(z)\rho(y) {\rm log}(\sqrt{z+c} + \sqrt{y+c})
- \frac{1}{12} c^3
\ea
where we used 
\be
\frac{\partial W}{\partial \lambda_c} = \frac{d}{dz}\frac{\delta W(\rho)}{\delta \rho(z)}
= w(z)
\ee
The parameter $c$ satisfies the saddle point equation for $w$ in (\ref{W}),
\be\label{c}
\frac{\partial W}{\partial c} = 
-\frac{1}{4}( c + \sum_d \frac{1}{\sqrt{\lambda_d + c}} )^2
 = 0
\ee
and this is consistent with (\ref{cs}).

The large P limit of the free energy $F$ in (\ref{Ft}) is obtained by the scaling
$\lambda_i \sim P^{2/3}$ and by taking each sum as order $P$. The order of $t_n$ becomes
\be
t_n = \sum_i \frac{1}{\lambda_i^{n + \frac{1}{2}}} \sim O(P^{\frac{2}{3}(1-n)})
\ee

The expansion of $c$ in (\ref{c}) is obtained by the recursive solution with the definition of
$t_n$ ($t_n = \sum \lambda_i^{-(\frac{1}{2}+n)}$),
\ba\label{c1}
c &=& - \sum_i \frac{1}{\sqrt{\lambda_i + c}}\nonumber\\
&=& - t_0 - \frac{1}{2} t_0 t_1 - \frac{3}{8} t_0^2 t_2 - \frac{1}{4} t_1^2 t_0 + 
O(\lambda^{-5})
\ea
From this equation, $c$ has to be negative, and $\lambda > - c$. Therefore, we have only
one expansion, the large $\Lambda$ expansion.

The free energy $W= {\rm log}Z$ in (\ref{W}) is divided into four terms. We expand 
each term for small $c$ ($c$ is a constant expressed by $t_n$),
\ba
W_1 &=& \frac{2}{3}\sum_i (\lambda_i + c)^{\frac{3}{2}}\nonumber\\
&=& \frac{2}{3}\sum_i \lambda_i^{\frac{3}{2}} + c \sum_i \lambda_i^{\frac{1}{2}}
+ \frac{c^2}{4}\sum_i \frac{1}{\lambda_i^{\frac{1}{2}}} 
- \frac{c^3}{24}\sum_i \frac{1}{\lambda_i^{\frac{3}{2}}} + \cdots
\nonumber\\
W_2 &=& - c \sum_i (\lambda_i + c)^{\frac{1}{2}}\nonumber\\
&=& - c \sum_i \lambda_i^{\frac{1}{2}} - \frac{c^2}{2}\sum_i \frac{1}{\lambda_i^{\frac{1}{2}}}
+ \frac{c^3}{8}\sum_i \frac{1}{\lambda_i^{\frac{3}{2}}}+ \cdots\nonumber\\
W_3 &=& - \frac{1}{2}\sum_{i,j} {\rm log} \bigg(\sqrt{\lambda_i + c} +\sqrt{\lambda_j + c}
\bigg)\nonumber\\
&=&- \frac{1}{2}\sum_{i,j}{\rm log}(\sqrt{\lambda_i}+\sqrt{\lambda_j}) - \frac{c}{4}
\sum_{i,j} \frac{1}{(\lambda_i\lambda_j)^{\frac{1}{2}}} + \frac{c^2}{8}
\sum_{i,j}  \frac{1}{\lambda_i^{\frac{1}{2}}\lambda_j^{\frac{3}{2}}}+\cdots
\nonumber\\
W_4 &=& - \frac{c^3}{12}
\ea
Inserting the expression of $c$, we obtain $W$, which is the sum of these four terms,
\be
W = \frac{2}{3}\sum_i \lambda_i^{\frac{3}{2}} - \frac{1}{2}\sum_{i,j}
 {\rm log}(\sqrt{\lambda_i}+\sqrt{\lambda_j})+ \frac{1}{12}t_0^3 + 
\frac{1}{24}t_0^3 t_1 + O(\frac{1}{\lambda^{\frac{9}{2}}})
\ee
The first two terms are ${\rm log} Z_0$ in (\ref{Z0}) and remainings are consistent with 
the genus zero part  of $F$ in (\ref{Ft}). Up to order $1/\lambda^3$, only  $t_0^3$ and $t_0^3 t_1$ are genus zero
terms.

The free energy $F$ is of order $P^2$ in the large P limit.
From (\ref{Ft}), we find in the large P limit,
\be\label{u1}
u = \frac{\partial^2 F}{\partial t_0^2} = \frac{1}{2}t_0 + \frac{1}{4}t_0 t_1 +
\frac{3}{16} t_0^2 t_2 + \frac{1}{8} t_0 t_1^2 + \cdots
\ee
Thus we find 
\be
c = - 2 u = - 2 \frac{\partial^2 F}{\partial t_0^2}
\ee

Therefore, we understand that $c$ is the specific heat for the free energy $F$, when we 
interprete 
$t_0$ as a temperature.

We now consider the Kontsevich model with a logarithmic term ($k\ne 0$).
The equations of motion in (\ref{virasoro}) are expressed as equations for $W$ and $W_a$, where

\be\label{W3}
\frac{\partial^3 Z}{\partial \lambda_c^3} = 
P \bigg(\frac{\partial^3 W}{\partial \lambda_c^3}\bigg)Z + 
3 P^2 \bigg( \frac{\partial^2 W}{\partial
\lambda_c^2}\bigg) W_c Z + P^3 W_c^3 Z
\ee

In the large P limit, the first and the second terms 
in (\ref{W3}) are negligible.

From (\ref{virasoro}), we express it as
\ba\label{w3} 
&&w^3(x) - x w(x) + 2 \int \frac{w(x) - w(u)}{(x-u)(u-v)} \rho(u)\rho(v) du dv\nonumber\\
&& - ( 1 + \frac{k}{P}) + \int du \frac{\rho(u)}{x-u} \bigg( 2 w^2(x) - w(x)w(u) - w^2(u) \bigg) = 0
\ea

From (\ref{w3}), we find in the large $x$ limit,
\be
w(x) \sim \sqrt{x} - \frac{1}{2 x}( 1 + \frac{k}{N}) 
\ee
This is a generalization of (\ref{asympto}) for $k\ne 0$.

The equation of (\ref{w3}) is a cubic equation. If $w(x)$ has a solution similar to
(\ref{wsol}), the tri-linear terms of $\rho$ has to be cancelled in this cubic equation of $w(x)$.
First we check that whether the solution of (\ref{wsol}) satisfies (\ref{w3}) when
$k=0$. We denote 
\be
l_x= \sqrt{x + c},\hskip 5mm l_y=\sqrt{y+c},\hskip 5mm l_z=\sqrt{z+c},\hskip 5mm
l_s = \sqrt{s+c}
\ee
The solution for $k=0$ is
\be
w(x) = l_x + \frac{1}{2}\int dy \frac{\rho(y)}{(l_x+l_y)l_y}
\ee
We express the cubic equation of (\ref{w3}) in terms of these $l_x,l_y,l_z,l_s$ by
\be
I_1+I_2+I_3+I_4+I_5 = 0
\ee
\ba
I_1&=& w^3(x) = l_x^{3}+ \frac{3}{2} l_x^2 \int dy \frac{ \rho(y)}{
(l_x+l_y)l_y}\nonumber\\
&+& \frac{4}{3}l_x \int  dy dz \frac{\rho(y)\rho(z)}{(l_x+l_y)(l_x+l_z)l_y l_z}\nonumber\\
&+& \frac{1}{8} \int dy dz ds \frac{\rho(y)\rho(z)\rho(s)}{(l_x+l_y)(l_x+l_z)(l_x+l_s)l_y 
l_z l_s}
\ea
\be
I_2= - x w(x)
\ee
\ba
I_3&=& 2 \int dy dz \frac{w(x)-w(y)}{(x-y)(y-z)} \rho(y)\rho(z)\nonumber\\
&=&2 \int dy dz \frac{l_x-l_y}{(x-y)(y-z)} \rho(y)\rho(z)\nonumber\\
&& + \int
dy dz ds \frac{\rho(y)\rho(z)\rho(s)}{(x-y)(y-z)l_s}(\frac{1}{l_x+l_s}-\frac{1}{l_y+l_s})
\ea
\be
I_4= -1
\ee
\ba
I_5= \int dy \frac{\rho(y)}{x-y}\bigg( 2 w^2(x)- w(x)w(y) -w^2(y)\bigg)
\ea
Up to the first order of $\rho$,
by adding the contribution $I$ of $I_1,I_2,I_3,I_4,I_5$, we have
\ba
\Delta I &=& \frac{c}{2}\int dy \frac{\rho(y)}{(l_x+l_y)l_y} \nonumber\\
&=& - \frac{1}{2}\int dy dz \frac{\rho(y)\rho(z)}{(l_x+l_y)l_y l_z}
\ea
where we have used the expression of $c$ given by (\ref{cs}). The summation of above
$\Delta I$ and the second order of $\rho$ in $I_1$ and $I_5$ becomes
\be
\Delta I + I_1 + I_5 = 2 \int dy dz \frac{\rho(y)\rho(z)}{(l_x+l_y)(z-y)}
\ee
and this is cancelled by the contribution of $I_3$.
Thus the contribution up to the second order is cancelled. The terms of the 
third order of $\rho$ come from $I_1$, $I_3$ and $I_5$. There are triple integrals
over $y$, $z$ and $s$. We symmetrize the integrals over these three 
variables.
Before making the  symmetrizations, we note that
\be
I_1= \frac{1}{8} \int dy dz ds \frac{\rho(y)\rho(z)\rho(s)}{(l_x+l_y)(l_x+l_z)(l_x+l_s)
l_y l_z l_s}
\ee
($I_1$ has symmetric form ).
\be
I_3= \frac{1}{2}\int dy dz ds \frac{\rho(y)\rho(z)\rho(s)(l_x+l_y+l_z+l_s)}{
(l_x+l_s)(l_x+l_y)(l_y+l_s)(l_x+l_z)(l_z+l_s)(l_y+l_z) l_s}
\ee
\be
I_5= -\frac{1}{4}\int dy dz ds \frac{\rho(y)\rho(z)\rho(s)( l_x+l_s +2l_y +2 l_z)}{
(l_x+l_y)(l_x+l_z)(l_y+l_z)(l_x+l_s)(l_y+l_s)l_z l_s}
\ee
After symmetrization, these three terms cancel completely $(I_1+I_3+I_5=0)$. 

Thus, we see that
the equation (\ref{w3}) is satisfied by the solution of $w(x)$  in (\ref{wsol}) when
$k = 0$. 

Since the parameter $k$ appears  only in terms of order zero of $\rho$ in
(\ref{w3}), it is  easily understood that
there is a straightforward solution for $k\ne 0$ case, based on the  above
analysis. Since we have seen the
solution of (\ref{wsol}) satisfies the integral equation of (\ref{w3}), we consider
the solution (\ref{wsol}) for $k\ne 0$ more carefully, specially the condition for
$c$. 
Namely, we use the same solution $w(x)$ as before
\be\label{wcx}
w(x) = \sqrt{x + c} + \frac{1}{2}\int_a^b dy \frac{\rho(y)}{(\sqrt{x+c}+\sqrt{y+c})\sqrt{y+c}}
\ee
where we consider that the parameter 
 $c$ is now a function of $x$, $c = c(x)$. We replace all parameters $c$ by $c(x)$ in
(\ref{wcx}). Since $x$ is fixed in the integral equation, this change from a constant
to x-dependence of $c$ does not make any difference.

Up to first order in $\rho$, by putting this 
$w(x)$ into the part of first order in $\rho$ of (\ref{w3}), we have
\ba\label{wk1}
 &&w^3(x) - x w(x) - (1 + \frac{k}{P}) + \int dy \frac{\rho(y)}{x-y}
\bigg(2 w^2(x) - w(x)w(y) - w^2(y)\bigg)\nonumber\\
&& = (x + c)^{\frac{3}{2}}- x \sqrt{x + c} - (1 + \frac{k}{P})
\nonumber\\
&&+ ( x + \frac{3}{2} c) \int_a^b \frac{\rho(y)}{(\sqrt{x+c}+\sqrt{y+c})\sqrt{y+c}}
\nonumber\\
&&+ \int_a^b dy \frac{\rho(y)}{x-y}\bigg(2 (x+c) -\sqrt{x+c}\sqrt{y+c} - (y+c)\bigg)
\nonumber\\
&=& c \sqrt{x + c} - \frac{k}{P} + \sqrt{x + c}\int_a^b dy \frac{\rho(y)}{\sqrt{y+c}}
+ \frac{c}{2} \int_a^b \frac{\rho(y)}{(\sqrt{x+c}+\sqrt{y+c})\sqrt{y+c}}\nonumber\\
\ea
where we have used that the integral of $\rho$ is one,
\be
\int_a^b \rho(y) dy = 1
\ee
We now put 
\be
c = - \int_a^b dy \frac{\rho(y)}{\sqrt{y+c}} + h
\ee
and r.h.s of (\ref{wk1}) becomes
\ba
&&r.h.s =h \sqrt{x+c} - \frac{k}{P} + \frac{h}{2} \int_a^b \frac{\rho(y)}{(\sqrt{x+c}+\sqrt{y+c})\sqrt{y+c}}
\nonumber\\
&&- \frac{1}{2}\int dz \frac{\rho(z)}{\sqrt{z+c}}\int_a^b \frac{\rho(y)}{(\sqrt{x+c}+\sqrt{y+c})\sqrt{y+c}}\nonumber\\
\ea
The last term is transfered to the part of  second order in $\rho$, and we have seen that the second 
and the third order terms of $\rho$ are cancelled completely. Therefore, we find that
\ba
&&h \bigg(\sqrt{x + c} + \frac{1}{2}\int_a^b \frac{\rho(y)}{(\sqrt{x+c}+\sqrt{y+c})\sqrt{y+c}}\bigg)
- \frac{k}{P}\nonumber\\
&&= h w(x) - \frac{k}{P} = 0
\ea
Thus we have  the coupled equations,
\ba\label{ck}
&&c = - \int_a^b dy \frac{\rho(y)}{\sqrt{y+c}} + \frac{k}{P} \frac{1}{w(x)},
\nonumber\\
&&w(x) = \sqrt{x + c} + \frac{1}{2} \int_a^b dy \frac{\rho(y)}{(\sqrt{x+c}+
\sqrt{y+c})\sqrt{y+c}}
\ea
This $w(x)$ satisfies the integral equation (\ref{w3}) and in the large $x$ limit,
it satisfies the asymptotic behavior
\be
w(x) \sim \sqrt{x} + \frac{1}{2 x}( 1 + \frac{k}{P}) \hskip 5mm (x \to \infty)
\ee

We obtain the expansion of $w(x)$ for large $x$ from the coupled equations in (\ref{ck}).
By  integration over $\rho(x)$  we obtain the free energy
$W$ for large $\lambda$ ($x= \lambda$),
\ba
W &=& \frac{2}{3}\sum \lambda_i^{\frac{3}{2}}+ \frac{k}{2}\sum {\rm log} \lambda_i - \frac{1}{2}
\sum_{i,j}{\rm log}(\sqrt{\lambda_i}+\sqrt{\lambda_j}) \nonumber\\
&+& 
\frac{1}{12}t_0^3  + \frac{1}{2} t_0 t_{\frac{1}{2}} + 
\frac{1}{4} k^2 t_1 + \cdots
\ea
which is consistent with ${\rm log} Z_0$ in (\ref{Z0}) and the genus zero part of 
$F$ in (\ref{Ft}).

When $k$ is sufficient large, we have a solution in which $c$ is positive in (\ref{ck}).
In this case, we obtain an expansion for small $\lambda$.

\vskip 5mm


\vskip 5mm
\section{Intersection numbers in a replica limit}
\setcounter{equation}{0}
\renewcommand{\theequation}{5.\arabic{equation}}

In the case of one matrix model, we have used a duality relation between the Kontsevich model
and the Gaussian random matrix model at a critical edge point \cite{BH1,BH2,BH3}. 
More precisely 
the Fourier transform of the n-point correlation function $U(s_1,...,s_n)$ becomes the
generating function of the intersection numbers with n-marked points. 
This n-point correlation function $U(s_1,...,s_n)$ has a Cauchy integral representation,
which is equivalent to the integral of the first Chern class over the moduli space
$\bar {\mathcal{M}}_{g,n}$.

We have shown in section 2, that there exists a similar duality relation between the partition function of the Kontsevich-Penner
model (\ref{intro1})
and the correlations for the Gaussian distribution (\ref{PP}). We want to discuss the origin of the
terms $t_{\frac{n}{2}}$ (half-integer) in this section.

In the expansion of the free energy $F$ in (\ref{Ft}), the number of times of appearance 
of $t_j$ is the number of
marked points according to the definition of the intersection number in (\ref{inter1}): 
In (\ref{inter1}), $n$ is the number of marked points.

We investigate first the case of one marked point (i.e. a single $t_j$). 
This case is obtained  from
the replica limit for  the matrix $B$ in (\ref{intro1}), namely the limit in which its size $P$  goes to zero. We first make
a shift $B \to B + \Lambda^{\frac{1}{2}}$ to eliminate the  linear term ${\rm tr} B \Lambda$,
\be\label{ZZ}
Z = \int_{P\times P} dB e^{-\frac{1}{3}{\rm tr} B^3 - {\rm tr} B^2 \Lambda^{\frac{1}{2}} 
+ 
k {\rm tr} {\rm log} (\Lambda^{\frac{1}{2}} + B)}
\ee
where $B$ is a $P\times P$ Hermitian matrix. The replica limit for $B$ means that we take
$P \to 0$ limit, selecting thereby  the contribution for one marked point \cite{BH2,BH4}. In this
replica limit, the number of eigenvalues $\lambda_i$ also goes to zero, and it becomes
not necessary to distinguish them. Therefore, we set $\Lambda = \lambda \cdot {\rm I}$.
Then, the calculation becomes considerably easier. First, we make the rescaling $B$ to $B/\lambda^{1/4}$.
Then the partition function $Z$ in (\ref{ZZ}) becomes
\be
Z = \int dB e^{- \frac{1}{3} \lambda^{-\frac{3}{2}}{\rm tr} B^3 - {\rm tr} B^2
+ k {\rm tr} {\rm log} (\lambda^{\frac{3}{4}} + B)}
\ee

We first neglect the cubic  vertex ${\rm tr} B^3$. 
Then, one recovers for  $Z$  the same model when $p=1$ in (\ref{intro2}), as was discussed 
in  references \cite{Chekhov,Kostov}. We consider  here this same model by the duality plus replica method.
\ba
Z &=& \int dB
e^{- {\rm tr} B^2 + k {\rm tr} {\rm log} (
\lambda^{\frac{3}{4}} + B )}\nonumber\\
&=& 2^{-\frac{P^2}{2}}\int_{P\times P} dB [{\rm det}( \lambda^{\frac{3}{4}} + 2^{-\frac{1}{2}} B )]^k e^{- \frac{1}{2}{\rm tr} B^2}
\ea

From the duality theorem for  characteristic polynomials  \cite{BH2}, the above 
expression has a dual form, which is an integral over   a $k\times k$ Hermitian matrix $M$ : 
\be
Z = \int_{k\times k} dM [{\rm det}( M + \lambda^{\frac{3}{4}} ) ]^P e^{-\frac{1}{2}{\rm tr} M^2}
\ee
 In the limit $P \to 0$, we have ($\hat \lambda = \sqrt{2} \lambda^{\frac{3}{4}}$)
\be\label{zetlambda}
\lim_{P\to 0} \frac{\partial Z}{\partial \hat \lambda} = \int_{k\times k} dM {\rm tr}\frac{1}{\hat \lambda
+ M} e^{-\frac{1}{2}{\rm tr} M^2}
\ee
which is the one-particle Green function $G(\hat \lambda)$ for the Gaussian random matrix, and
the expansion of the inverse of $\hat \lambda$ is easily obtained as  a moment of $M$ 
in terms of polynomials of $k$.

By  integration of $G(\hat \lambda)$ about $\hat \lambda$, we obtain $Z$,
\be\label{GG}
Z = k {\rm log} \hat \lambda + \sum_{j=1}^\infty \frac{1}{(2 j)\hat \lambda^{2 j}}< {\rm tr}M^{2 j} >
\ee

The Gaussian average $<{\rm tr} M^{2j} >$ is easily evaluated from the integral representation of
$U(s)$ \cite{BH2},
\ba
U(s) &=& <{\rm tr} e^{s M} >\nonumber\\
&=&\frac{1}{s}e^{\frac{1}{2}s^2}\oint \frac{du}{2 i \pi} (1 + \frac{s}{u})^k e^{s u}
\ea
This integral runs over  the contour centered at  $u=0$, and it yields
\be
U(s) = k + \frac{k^2}{2}s^2 + \frac{2 k^3 + k}{24} s^4 + \frac{k^4 + 2 k^2}{144} s^6 + 
\frac{2 k^5 + 10 k^3 + 3 k}{5760} s^8 + \cdots
\ee
From these expressions, we obtain
\ba
&&<{\rm tr}M^2> = k^2,\hskip 2mm <{\rm tr}M^4>= 2 k^3+k,\hskip 2mm <{\rm tr} M^6> = 5 k^4 + 10 k^2,
\nonumber\\
&&<{\rm tr}M^8> = 14 k^5 + 70 k^3 + 21 k, \hskip 2mm \cdots
\ea

From (\ref{GG}) , we express $Z$ of (\ref{GG}) in terms of $t_j$ ($t_j= \frac{1}{ \lambda^{j + \frac{1}{2}}}$),
by noting that $s = \frac{1}{\hat \lambda} = - \frac{1}{\sqrt{2}\lambda^{\frac{3}{4}}}$,
\be\label{log2}
Z = k {\rm log}\hat \lambda + \frac{k^2}{4}t_1 + \frac{1}{16}( k + 2 k^3) t_{\frac{5}{2}} + \frac{1}{48}
(5 k^4+10 k^2) t_{4}+ \cdots
\ee
The term $\frac{1}{4} k^2 t_1$ coincides with the term in $F$ of (\ref{Ft}).

The tri-valent term, which we have neglected, couples to the logarithmic term, and also 
make a contribution as polynomials in $k$. The exponent 
${\rm exp}(- \frac{1}{3} \lambda^{-\frac{3}{2}}{\rm tr} B^3)$ is expanded and it gives the
contribution in the replica limit $P\to 0$.
By the formula of the replica limit, we have nonvanishing average of
 $<\prod_i {\rm tr } B^{d_i}>$. This formula is \cite{BH2}
\ba\label{replica}
\lim_{P\to 0}U(s_1,...,s_l)&=& \lim_{P\to 0}
\frac{1}{P}< {\rm tr}e^{s_1 B}\cdots {\rm tr}e^{s_l B}>\nonumber\\
&=&\frac{1}{\sigma^2}\prod_{j=1}^l 2{\rm sinh}\frac{s_j \sigma}{2}
\ea
where $\sigma = s_1+ \cdots + s_l$. This provides  a generating function for
$<\prod_i {\rm tr } B^{d_i}
>$.
 
From this formula, for instance, we have
\be
\lim_{P\to 0}\frac{1}{P}< {\rm tr} B^3 {\rm tr} B^3 > = 3 , \hskip 5mm 
\lim_{P\to 0}\frac{1}{P}<{\rm tr B^3}{\rm tr} B^3 {\rm tr}B^2> = 18,
\cdots
\ee

Using these values of averages, we are in position  to compute  the coefficients of  the $t_n$ terms.
We consider the term $t_{\frac{5}{2}}$ in (\ref{Ft}), which has a coefficient
$\frac{2}{3}(k + k^3)$. 
The partition function $Z$ is
\be
Z = \int dB e^{-\frac{1}{2}{\rm tr} B^3 - {\rm tr} B^2 \Lambda^{\frac{1}{2}} + k {\rm tr}{\rm log}
(1 + \Lambda^{-\frac{1}{2}}B)}
\ee
We rescale $B\to 2^{-\frac{1}{3}}\lambda^{-\frac{1}{4}}B$ and $\lambda^{\frac{1}{2}}\to 2^{-\frac{1}{3}}\lambda^{\frac{1}{2}}$.
We have
\be
Z = \int dB e^{-\frac{1}{6 \lambda^{3/2}}{\rm tr} B^3 - \frac{1}{2}{\rm tr}B^2 + k {\rm tr}(1 + \lambda^{-\frac{3}{4}}
B)}
\ee
Expanding then  the logarithmic term and ${\rm exp}(-\frac{1}{6 \lambda^{3/2}}{\rm tr} B^3 )$ term,
we find the  contributions to order $\frac{1}{\lambda^3}$ from 6 terms. These terms  are evaluated by the replica formula
 ($P\to 0$ limit) of  (\ref{replica}),
\ba
&&(i)\frac{1}{P}< \frac{k}{4\lambda^3}{\rm tr}B^4> = \frac{k}{4 \lambda^3},\nonumber\\
&&(ii)\frac{1}{P} < \frac{1}{6\lambda^{\frac{3}{4}}}{\rm tr} B^3\cdot 
\frac{1}{3!}(\frac{k}{\lambda^{\frac{3}{4}}})^3
({\rm tr}B)^3> = \frac{k^3}{6 \lambda^3}\nonumber\\
&&(iii)\frac{1}{P}< \frac{k}{2}(\frac{1}{\lambda^{\frac{3}{4}}})^2 {\rm tr}B^2\cdot 
\frac{1}{2!}(\frac{k}{\lambda^{\frac{3}{4}}})^2 ({\rm tr}B)^2> \frac{k^3}{2\lambda^3}\nonumber\\
&&(iv) \frac{1}{P}<\frac{1}{6 \lambda^{\frac{3}{4}}}{\rm tr}B^3\cdot \frac{k}{3}(\frac{1}{\lambda^{\frac{3}{4}}})^3
{\rm tr}B^3> = \frac{k}{6 \lambda^{\frac{3}{4}}}\nonumber\\
&&(v)\frac{1}{P}<\frac{1}{2}(\frac{1}{6\lambda^{\frac{3}{4}}})^2 ({\rm tr}B^3)^2\cdot \frac{k}{2}(\frac{1}{
\lambda^{\frac{3}{4}}})^2 {\rm tr} B^2> = \frac{k}{8 \lambda^3}\nonumber\\
&&(vi)\frac{1}{P}<\frac{k}{\lambda^{\frac{3}{4}}}{\rm tr} B\cdot \frac{1}{3!} (\frac{1}{6\lambda^{\frac{3}{4}}}
{\rm tr} B^3)^3> = \frac{k}{8 \lambda^3}
\ea
Adding these (i)$\sim$(vi) terms, and noting that we have made a scaling of $\lambda$, we obtain
as expected  the result $\frac{1}{6}(k^3+ k)\lambda^{-3} = \frac{1}{6}(k^3+k)t_{\frac{5}{2}}$  for (\ref{Ft}).

We now evaluate the coefficients of $t_j$ from the Fourier transform of one-point correlation function $U(s)$.
We have shown that the intersection numbers of n-marked point are obatined from the Fourier transforms
of n-point correlation function $U(s_1,...,s_n)$ \cite{BH3,BH4}.  The Kontsevich-Penner model involves two  interaction terms
${\rm tr} B^3$ and ${\rm tr} {\rm log}B$. For the application of $U(s_1,...,s_n)$  to  this Kontsevich-Penner model,
we have  to extend the previous duality expression. 

  Let us return to the  duality relation for the one-matrix model, before extending it ; in this one-matrix case the duality reads
\be\label{duality1}
< \prod_{\alpha=1}^k {\rm det}(\lambda_\alpha- M)>_{M,A} = <\prod_{j=1}^N {\rm det}(a_j - i B)>_{B,\Lambda}
\ee
in which the l.h.s. consists of Gaussian average in an external matrix source $A$ for Hermitian $N\times N$ matrices ; the r.h.s. is also a Gaussian average for Hermitian $k\times k$ matrices in an external source $\Lambda$ (whose eigenvalues are the $\lambda_{\alpha}$ of the l.h.s. ; the $a_j$ are the eigenvalues of $A$). 
We also know in closed form the Fourier transform of the n-point correlation function,
\ba
&&U(s_1,...,s_n) = < \prod_{j=1}^n {\rm tr} e^{s_j M} >_{M,A}\nonumber\\
&&= \oint \prod \frac{dU_i}{2 i \pi} e^{N \sum u_l s_l+ \frac{1}{2}N \sum s_l^2}
\prod_{l=1}^n \prod_{j=1}^N ( 1 + \frac{s_l}{u_l-a_j}){\rm det}\frac{1}{u_i-u_j+ s_i}
\ea
Let us consider the one-point function  
\ba\label{us1}
U(s)& = &\frac{1}{s} \oint \frac{du}{2 i \pi}\prod_{j=1}^N ( 1 + \frac{s}{u-a_j}) {\rm exp}( N u s
+ \frac{1}{2}N s^2) \nonumber\\
&=& \frac{1}{s} \oint \frac{du}{2 i \pi} {\rm exp}(-\sum_{m=1}^\infty c_m [(u + s)^m - u^m] + N u s
+ \frac{1}{2}N s^2)
\ea
where
\be
c_m = \frac{1}{m} \sum_{j=1}^N \frac{1}{(a_j)^m}
\ee

The r.h.s. of the duality formula  (\ref{duality1}) is also expanded in powers of $B$,
\be
< \prod {\rm det}(a_j - i B)> = \int dB
{\rm exp}(- \sum c_m (i B)^m - \frac{1}{2}N {\rm tr} B^2 + N {\rm tr} B \Lambda)
\ee

From this representation we have investigated the (p,1)-model (the (2,1) corresponds to Kontsevich model),
 which is obtained by specifying appropriately the  $a_j$ \cite{BH3}.
We  consider here a more general situation, in view of encompassing   the Kontsevich-Penner model.
The  (p,q)-model  is defined  by
\be\label{pqm}
Z = \int dB {\rm exp}( - c_{p+1} {\rm tr} B^{p+1} -  c_{q+1} {\rm tr} B^{q+1} + {\rm tr} B\Lambda )
\ee
This is obtained by imposing the following conditions to the $a_j$,
\ba
    && \frac{1}{2}\sum_{j=1}^N \frac{1}{(a_j)^2} = \frac{N}{2}\nonumber\\
&& \frac{1}{m}\sum_{j=1}^N \frac{1}{(a_j)^m} = 0
\ea
where $m= 3,4,...,q$ and $m \ne p+1$.
These conditions should be understood as holding in the large N limit.

These conditions may also be  applied to $U(s)$ in (\ref{us1}).  Then we have
\be
U(s) = \frac{1}{s}\oint \frac{du}{2 i \pi} e^{- c_{q+1}((u+s)^{q+1}-u^{q+1})- c_{p+1}((u+s)^{p+1}-u^{p+1})}
\ee
For the application to the Kontsevich-Penner model, we have to take the limit $p\to -1$ and $q\to 2$. Since $c_{p+1}
= \frac{1}{p+1}\sum \frac{1}{(a_j)^{p+1}}$.
we have
\be
Z = \int_{k\times k} dB {\rm exp}(-c_3 {\rm tr} B^3 - N {\rm tr}{\rm log} B + {\rm tr}B \Lambda)
\ee
and
\be
U(s) = \frac{1}{s}\oint \frac{du}{2 i \pi} e^{- c_3 ((u+s)^3-u^3)- N {\rm tr}{\rm log}(\frac{u+s}{u})}
\ee
For the n-point correlation functions,
we have similarly
\be\label{npointU}
U(s_1,...,s_n) = \oint \prod_{i=1}^n
\frac{du_i}{2 i \pi} e^{- \sum_i c_3 ((u_i + s_i)^3 - u_i^3) - N \sum_i {\rm tr}{\rm log}(\frac{u_i+s_i}{
u_i})}{\rm det}\frac{1}{u_i-u_j+s_i}
\ee

The intersection numbers for one-marked point, which are obtained from $Z$ as the coefficients of the linear terms in the $t_j$
\be
Z = \sum <\tau_j> t_j + {\rm{ higher \  degree}},
\ee
are derived from $U(s)$ as coefficients of the expansion in powers of $s$ .

Apart from notations, in which we have to interchange $N$ to $k$, and further  $k\to -k$,  and 
chose $c_3 = \frac{1}{3}$,  $Z$ is then identical to  the
Kontsevich-Penner model in (\ref{intro1}).
We then have
\be
U(s) = \frac{e^{-\frac{1}{3}s^3}}{s}\oint \frac{du}{2 i \pi} e^{- s u^2 - s^2 u} (\frac{u+s}{u})^k
\ee
We now expand  $U(s)$ in powers of $s$ andl $k$. To that purpose we shift 
 $u= v-\frac{1}{2}s$, and $v= s z/2$, and obtain
\be\label{usx}
U(s) = \frac{1}{2} e^{-\frac{1}{12} s^3} \int \frac{dz}{2 i \pi} e^{-\frac{1}{4} s^3 z^2}(\frac{z+1}{z-1})^k
\ee
This gives an expansion in powers of  $k$, from
\ba
(\frac{z+1}{z-1})^k &=& 1 + k {\rm log}(\frac{z+1}{z-1}) + \frac{1}{2} k^2 [{\rm log}(\frac{z+1}{z-1})]^2
\nonumber\\
&+& \frac{k^3}{6}[{\rm log}(\frac{z+1}{z-1})]^3 + O(k^4)
\ea
The first order of $k$ leads to the integral
\be
\int_{-\infty}^\infty dz e^{-\frac{1}{4}s^3 z^2} {\rm log}(\frac{z+1}{z-1}) = 
2 i (\frac{\pi}{s})^{\frac{3}{2}} erf(\frac{1}{2}s^{\frac{3}{2}})
\ee
where $erf(x)$ is the error function. For small $s$ and small $k$, we obtain
\be\label{s3k1}
U(s) = k - \frac{1}{6} k s^3 + O(k s^6)
\ee
The integrals for odd powers of the logarithm may be computed analytically. For instance, from (\ref{usx}), in the the  small $s$ and $k$ expansion, we obtain a term of order of $s^3 k^3$,
\ba\label{s3k3}
I &=& - \frac{s^3 k^3}{2} \int_{-\infty}^\infty dz (\frac{1}{12} + \frac{z^2}{4}) \frac{1}{6}
 [{\rm log}(\frac{z+1}{z-1})]^3\nonumber\\
&=& \frac{1}{6} i \pi s^3 k^3
\ea
Therefore we recover from (\ref{s3k1}) and (\ref{s3k3}), the coefficient of $s^3$ as
$\frac{1}{6} ( k + k^3) s^3$, which yeilds the expected $\frac{1}{6}(k + k^3) t_{\frac{5}{2}}$ term in (\ref{Ft}).

We note that the odd powers of $k$ may be obtained systematically by computing 
a residue  at  $z=1$ in the contour integral (\ref{usx}). For the even powers of $k$, we have to integrate the logarithmic terms.

The replica formula  (\ref{replica}) has been derived with a  single logarithmic
integral, which has a cut for $-s < u < 0$ \cite{BH2}. This replica formula corresponds to the $k\to 0$ 
limit of the(p,q)-model with $q=1$ and $p=-1$, which leads to the intersection numbers with one-marked point.
In an appendix, we present the two-point function $U(s_1,s_2)$ for $q=1$ and $p=-1$ as a polynomial in $k$, consistent
with the replica formula for $k \to 0$. From this example, we understand that the contours for the n-point 
function should encompass all poles. We now consider the  two-point case for the Kontsevich-Penner model with two marked points : 
\be
U(s_1,s_2) = e^{\frac{1}{3}(s_1^3+s_2^3)}\oint \frac{du_1 du_2}{(2 i \pi)^2} \frac{e^{u1^2 s_1+ s_1^2 u_1+ u_2^2 s_2
+ u_2 s_2^2}}{(u_1-u_2+ s_1)(u_1- u_2+s_2)} (\frac{u_1+s_1}{u_1})^k (\frac{u_2+ s_2}{u_2})^k
\ee
where the contours for $u_2$ are such that one sums over the three contributions from the poles at  $u_2=0,u_2=u_1+s_1, u_2= u_1+s_2$ and then the contour for  $u_1$ circles around the origin.  This yields
\ba
U(s_1,s_2) &=& k^2 (s_1^2 s_2 + s_1 s_2^2) + \frac{k^2}{12}(s_1^5 s_2+ s_1 s_2^5) \nonumber\\
&+& (\frac{k^4}{4}+ \frac{7}{12} k^2)(s_1^4 s_2^2+ s_1^2 s_2^4) +
(\frac{k^4}{4} + \frac{3}{4} k^2) s_1^3 s_2^3 + O(s^9)
\ea
Changing from $s^m$ to $t_{m-\frac{1}{2}}$, we obtain the terms with two marked points of (\ref{Ft}).

Thus we find that the terms with n-marked point for the Kontsevich-Penner model (\ref{intro1}) are expressed 
explicitly by the integral formula of $U(s_1,...,s_n)$ of (\ref{npointU}). The Kontsevich-Penner model
is the  $p=-1,q=2$ of the (p,q)-model of (\ref{pqm}). Following the same techniques as above we find similarly
 explicit integral representation of $U(s_1,...,s_n)$ for the (p,q)-model with $p=-1$ and arbitary $q$.

\vskip 5mm
\section{Discussions}
\setcounter{equation}{0}
\renewcommand{\theequation}{6.\arabic{equation}}

In  this paper, we have extended the analysis of our previous paper \cite{BH4} to
the Kontsevich-Penner model. We have derived the Virasoro constraints for this model, and 
we have obtained the large N solution of the corresponding integral equation.
The occurence of  parameters $t_{n}$, with half-integer $n$, is due to the  logarithmic potential. 
Using  the correlation functions of the  Gaussian two-matrix model, in a source, with one set of eigenvalues near an edge, and the other
one in the bulk of the spectrum, provides the Kontsevich-Penner model.

We have  used an explicit integral representation for $U(s_1,...,s_n)$  (\ref{npointU}), which gives
then $k$-dependent coefficients of the free energy $F$ of the Kontsevich-Penner model. This
model turns out to be the special limit  $p=-1$ and $q=2$  of a $ (p,q)$-model . The integral representation for $U(s_1,...,s_n)$ 
is valid also for general $q$ with $p=-1$. The details for such cases is left to future work.

In string theory, the $c=1$ matrix model has attracted considerable interest, renewed recently
from the D-brane point of view. The tachyon plays a central role in the $c=1$ matrix model. In the present study, the
partition function for the Kontsevich-Penner model (\ref{intro1}),  is derived from a time dependent
Gaussian matrix model, and the role of the $t_{\frac{n}{2}}$ and $k$-dependence are clearly understood
from the correlation functions of  the two-matrix model. Thus, it may shed a light on the $c=1$ string
theory, FZZT-brane, etc \cite{Mukhi3}.
\vskip 5mm
{\bf{Acknowledgement}}
S.H. is supported by a Grant-in Aid for Scientific Research (C) of JSPS.

\newpage
\begin{center}
{\bf Appendix A: formula for $p$-th derivatives}
\end{center}
\setcounter{equation}{0}
\renewcommand{\theequation}{A.\arabic{equation}}
\vskip 5mm
The matrix $\Lambda$ has eigenvalues $\lambda_1,\lambda_2,....$ and corresponding orthonormal eigenfunctions $\vert \phi_a\rangle$. 
We now consider a perturbation  $d \Lambda$, 
\be
(\Lambda + d \Lambda )\bigg( |\phi> + |d\phi> \bigg) = (\lambda + d \lambda)
\bigg( |\phi> + |d\phi> \bigg)
\ee
and from this equation, we obtain at first order
\be\label{perturb1}
(\Lambda - \lambda_a)|d\phi_a> + (d\Lambda - d \lambda_a)|\phi_a> = 0
\ee
Multiplying $<\phi_a|$ from the left side, it becomes
\be
<\phi_a| d \Lambda |\phi_a> = d \lambda_a
\ee
In an arbitrary fixed orthonormal basis $|b>$, it becomes
\be
d\lambda_a = <\phi_a|b><b|d\Lambda|c><c|\phi_a>
\ee
Therefore, we obtain the first important formula,
\be\label{formula1}
\frac{\partial \lambda_a}{\partial \Lambda_{bc}} = <\phi_a|b><c|\phi_a>
\ee
Note that $<\phi_a|b>= U_{ab}$, where $U$ is a unitary matrix.
From (\ref{perturb1}),  multiplying  by $<\phi_b|$  ($b\neq a$) the left hand side,
\be
<\phi_b|d\phi_a> = \frac{1}{\lambda_a - \lambda_b} <\phi_b|d\Lambda |\phi_a>
\ee
Therefore, we have
\be
 |d\phi_a> = \sum_{b\neq a} \frac{1}{\lambda_a -\lambda_b}|\phi_b>
<\phi_b|d\Lambda|\phi_a>
\ee

from which follows the second formula,
\be\label{formula2}
\frac{\partial <b|\phi_a>}{\partial \Lambda_{cd}} = 
\sum_{f\ne a} \frac{1}{\lambda_a- \lambda_f}<b|\phi_f><\phi_f|c><d|\phi_a>
\ee
The conjugate of this formula is
\be
\frac{\partial <\phi_a|b>}{\partial \Lambda_{dc}} = 
\sum_{f\ne a} \frac{1}{\lambda_a-\lambda_f}<\phi_f|b><c|\phi_f><\phi_a|d>
\ee
By the chain rule, we obtain the first derivative,
\ba
\frac{\partial Z}{\partial \Lambda_{ab}} &=&
 \frac{\partial \lambda_c}{\partial \Lambda_{ab}}
\frac{\partial Z}{\partial \lambda_c}\nonumber\\
 &=& <b|\phi_c><\phi_c|a> \bigg(\frac{\partial Z}{\partial \lambda_c}\bigg)
\ea

The formula for  the second derivative is obtained by the use of (\ref{formula1}) and (\ref{formula2}).
\ba
\bigg(\frac{\partial^2}{\partial \Lambda^2 }\bigg)_{ab} Z &=& \bigg(\frac{\partial}{\partial \Lambda}\bigg)_{ad}\bigg(\frac{\partial}{
\partial \Lambda}\bigg)_{db} Z\nonumber\\
&=& \frac{\partial}{\partial \Lambda_{ad}}\bigg( <b|\phi_c><\phi_c|d>\bigg(\frac{\partial Z}{\partial \lambda_c}\bigg)
\bigg)
\ea
Noting that
\ba
&&<\phi_c|d>\bigg(\frac{\partial Z}{\partial \lambda_c}\bigg) 
\frac{\partial}{\partial \Lambda_{ad}}<b|\phi_c>\nonumber\\
&&= <\phi_c|d>\bigg(\frac{\partial Z}{\partial \lambda_c}\bigg)\sum_f \frac{1}{\lambda_c-\lambda_f}<b|\phi_f><\phi_f|a><d|\phi_c>
\nonumber\\
&& = \sum_d <b|\phi_c><\phi_c|a>\bigg(\frac{\partial Z}{\partial \lambda_d}\bigg) 
\frac{1}{\lambda_d - \lambda_c}
\ea
and
\ba
&&<b|\phi_c>\bigg(\frac{\partial Z}{\partial \lambda_c}\bigg) \frac{\partial }{\partial \Lambda_{ad}}<\phi_c|d>\nonumber\\
&&= <b|\phi_c>\bigg(\frac{\partial Z}{\partial \lambda_c}\bigg)
\sum_f \frac{1}{\lambda_c- \lambda_f} <\phi_f|d><d|\phi_f><\phi_c|a>\nonumber\\
&&= <b|\phi_c><\phi_c|a>\bigg(\frac{\partial Z}{\partial \lambda_c}\bigg)\sum_d 
\frac{1}{\lambda_c - \lambda_d}
\ea
we obtain
\be
\bigg(\frac{\partial^2}{\partial \Lambda^2}\bigg)_{ab} Z = 
<b|\phi_c>\bigg( \frac{\partial^2}{\partial {\lambda_c}^2}
+ \sum_{d\ne c} \frac{1}{\lambda_c -\lambda_d} \bigg(\frac{\partial Z}{\partial \lambda_c}- \frac{\partial Z}{\partial \lambda_d}
\bigg) \bigg) <\phi_c|a>
\ee

The third order differentiation is obtained by repeating the same procedure.
We write $\Gamma_c$ by
\be
\Gamma_c = \frac{\partial^2 }{\partial {\lambda_c}^2} 
+ \sum_{d\ne c}\frac{1}{\lambda_c-\lambda_d}
\bigg(\frac{\partial }{\partial \lambda_c}-\frac{\partial }{\partial \lambda_d}\bigg)
\ee
 
\ba
&&\bigg(\frac{\partial^3}{\partial \Lambda^3}\bigg)_{pb} 
= \bigg(\frac{\partial}{\partial \Lambda}\bigg)_{pa}\bigg(
\frac{\partial^2}{\partial \Lambda^2}\bigg)_{ab} \nonumber\\
&=& \bigg(\frac{\partial \lambda_c}{\partial \Lambda_{pa}}\bigg)
\frac{\partial}{\partial \lambda_c}
\bigg( <b|\phi_c> \Gamma_c <\phi_c|a>\bigg)\nonumber\\
&=& <b|\phi_c><\phi_c|p><a|\phi_c><\phi_c|a>\frac{\partial \Gamma_c}{\partial \lambda_c}
\nonumber\\
&&+ <\phi_c|p><a|\phi_c>\Gamma_c <\phi_c|a>\frac{\partial <b|\phi_c>}{\partial \lambda_c}
\nonumber\\
&&+ <\phi_c|p><a|\phi_c><b|\phi_c>\Gamma_c \frac{\partial <\phi_c|a>}{\partial \lambda_c}
\ea
Therefore, we obtain
\be
\bigg(\frac{\partial^3}{\partial \Lambda^3}\bigg)_{ab}  = <b|\phi_c>\bigg(
\frac{\partial \Gamma_c}{\partial \lambda_c} + \sum_{d\ne c} \frac{1}{\lambda_c-\lambda_d}(
\Gamma_c-\Gamma_d)\bigg) <\phi_c|a>
\ee
Using the identity,
\be
\frac{1}{(\lambda_c-\lambda_d)(\lambda_c-\lambda_e)}
+\frac{1}{(\lambda_d-\lambda_c)(\lambda_d-\lambda_e)}
+\frac{1}{(\lambda_e-\lambda_c)(\lambda_e-\lambda_d)}=0
\ee
we obtain the expression in terms of eigenvalues 
\ba
 &&\bigg(\frac{\partial^3}{\partial \Lambda^3}\bigg)_{ab} 
=
\frac{\partial^3 }{\partial {\lambda_c}^3} \nonumber\\
&&+ \sum_{d\ne c} \frac{1}{\lambda_c-\lambda_d}
(\frac{\partial}{\partial \lambda_c} - \frac{\partial}{\partial \lambda_d})
(2 \frac{\partial}{\partial \lambda_c} + \frac{\partial}{\partial \lambda_d}) 
- \sum_{d\ne c}\frac{1}{(\lambda_c- \lambda_d)^2}
(\frac{\partial}{\partial \lambda_c}-\frac{
\partial}{\partial \lambda_d})\nonumber\\
&&+ 2 \sum_{d\ne e,c}\sum_{e\ne c} \frac{1}{(\lambda_c -\lambda_e)(\lambda_e - \lambda_d)}(\frac{\partial}{
\partial \lambda_c}-\frac{\partial}{\partial \lambda_e})
\ea

If we write 
\be
\Gamma_c^{(1)} = \frac{\partial }{\partial \lambda_c}
\ee
we have
\be
\bigg(\frac{\partial^2}{\partial \Lambda^2}\bigg)_{ab} = <b|\phi_c>\Gamma_c^{(2)}<\phi_c|a>
\ee
\be
\Gamma_c^{(2)} = \frac{\partial}{\partial \lambda_c} \Gamma_c^{(1)}+ \sum_d \frac{1}{\lambda_c-
\lambda_d}(\Gamma_c^{(1)}-\Gamma_d^{(1)})
\ee
Repeating this procedure, we obtain
\be
\Gamma_c^{(p+1)} = \frac{\partial}{\partial \lambda_c} \Gamma_c^{(p)}+ \sum_d \frac{1}{\lambda_c-
\lambda_d}(\Gamma_c^{(p)}-\Gamma_d^{(p)})
\ee
and
\be
\bigg(\frac{\partial^{p+1}}{\partial \Lambda^{p+1}}\bigg)_{ab} = <b|\phi_c>\Gamma_c^{(p+1)}<\phi_c|a>
\ee

\vskip 5mm
\begin{center}
{\bf Appendix B: Relation to unitary matrix model }
\end{center}
\setcounter{equation}{0}
\renewcommand{\theequation}{B.\arabic{equation}}
\vskip 3mm
 We will show that the unitary matrix model with
external source \cite{BG} is equivalent to the higher Airy matrix model
with a logarithmic potential for $p= -2$ \cite{BH4a,Mironov}.

The unitary matrix model (Brezin-Gross model \cite{BG}) is
\be\label{BG}
Z = \int dU e^{{\rm tr} (U A^{\dagger} + U^{\dagger} A)}
\ee
where $U$ is a unitary matrix and $A$ is an arbitrary complex matrix.
From the unitarity condition, $U U^{\dagger}=1$, we have
\be
\frac{\partial^2}{\partial A^{\dagger} \partial A} Z = I \cdot Z
\ee
Introducing $\Lambda$ as
\be
\Lambda = A A^{\dagger} \hskip 5mm (A_{ij} A_{jk}^{\dagger} = \Lambda_{ik})
\ee
we find
\ba
&&\frac{\partial^2}{\partial A_{ij}^{\dagger} \partial A_{jk}} =
\frac{\partial}{\partial A_{ij}^{\dagger}} \frac{\partial \Lambda_{qs}}{\partial A_{jk}}
\frac{\partial }{\partial \Lambda_{qs}}\nonumber\\
&&= \frac{\partial}{\partial \Lambda_{ks}} \Lambda_{ms}\frac{\partial}{\partial \Lambda_{mi}}
\nonumber\\
&=& \Lambda \frac{\partial^2}{\partial \Lambda^2} + N \frac{\partial}{\partial \Lambda}
\ea
Thus
\be\label{uu1}
\bigg(\Lambda \frac{\partial^2}{\partial \Lambda^2} + N \frac{\partial}{\partial \Lambda}
\bigg) Z = I\cdot Z
\ee
The equations of motion for the $p=-2$ case follows from 
\be
\int dM \frac{\partial}{\partial M} e^{{\rm tr} M^{-1} + {\rm tr} M \Lambda + k {\rm tr}
{\rm log} M } = 0
\ee
Taking two derivatives with respect to $\lambda$, this leads to
\be\label{uu2}
\bigg( -1 + 2 N \frac{\partial}{\partial \Lambda} + 
\Lambda \frac{\partial^2}{\partial \Lambda^2}
+ k \frac{\partial}{\partial \Lambda} \bigg) Z = 0
\ee
If we now  take $k=-N$, (\ref{uu2}) this equation becomes identical to  (\ref{uu1}).

Thus we find that unitary matrix model with  an external source is  similar
to the Kontsevich model with a logarithmic potential.
There is a phase transition  in this unitary matrix model with a critical point at 
\be
s = {\rm tr} \frac{1}{\sqrt{A A^{\dagger}}} = \sum_{i=1}^N
 \frac{1}{\sqrt{\lambda_i}}= 2.
\ee

\newpage


\begin{thebibliography}{99}

\bibitem{BH1}
   E. Br\'ezin and S. Hikami, Vertices from replica in a random matrix theory, 
J. Phys. A.  40 (2007) 13545. arXiv:0704.2044[math-ph].
\bibitem{BH2}
   E. Br\'ezin and S. Hikami, Intersection theory from duality and replica. 
Comm. Math. Phys. 283 (2008) 507. arXiv:0708.2210[hep-th].
\bibitem{BH3}
   E. Br\'ezin and S. Hikami, Intersection numbers of Riemann surfaces from 
Gaussian matrix models. JHEP 10 (2007) 096, arXiv:0709.3378. 
\bibitem{BH4}
   E. Br\'ezin and S. Hikami, Computing topological invariants with one and two-matrix models,
JHEP 04 (2009) 110, arXiv:0810.1085.
\bibitem{BH4a}  E. Br\'ezin and S. Hikami, Duality and Replicas for a unitary matrix model,
JHEP 07 (2010) 067, arXiv:1005.4730.
\bibitem{Witten1}
  E. Witten, Algebraic geometry associated with matrix models of two dimensional gravity,
in "Topological Methods in Modern Mathematics", Publish or Perish INC. 1993, New York. P.235.
\bibitem{Witten2}
  E. Witten, On the Kontsevich model and other models of two dimensional gravity, IASSNS-HEP-91/24,
  (1991).
\bibitem{Kontsevich}
    M. Kontsevich, Intersection theory on the moduli space of curves and the matrix Airy function,
 Commun. Math. Phys. 147, 1 (1992).
\bibitem{B}
   E. Br\'ezin, V. Kazakov and Al.B. Zamolodchikov, Scaling violation in a field theory of 
closed strings in one physical dimension,
 Nucl. Phys. B338, 673 (1990).
\bibitem{Mironov}
   A. Mironov, A. Morozov and G.W.Semenoff, 
Unitary matrix integrals in the framework of Generalized Kontsevich Model
1. Br\'ezin-Gross-Witten model. arXiv:hep-th/9404005.
\bibitem{BG}
E. Br\'ezin and D.J. Gross, 
The external field problem in the large N limit of QCD.
Phys. Lett. B97 (1980) 120.
\bibitem{GW}
D. Gross and E. Witten, Phys. Rev. D21 (1980) 446.
Possible third-order phase transition in the large-N lattice gauge
theory.
\bibitem{Chekhov}
  L. Chekhov and Yu. Makeenko, A hint on the external field problem for matrix models,
arXiv: hep-th/9202006.
\bibitem{Kostov}
    I. K. Kostov and M. L. Mehta, Phys. Lett. 189B (1987) 118.
\bibitem{Mukhi1}
   C. Imbimbo and S. Mukhi, The topological matrix model of c=1 string, Nucl. Phys. B449,553 (1995).
\bibitem{Mukhi2}
   S. Mukhi, Topological matrix models, Liouville matrix model and c=1 string theory,
arXiv:hep-th/0310287.
\bibitem{Ambjorn}
J. Ambjorn and L. Chekhov, The NBI matrix model of IIB superstrings, JHEP 9812:007(1998).
\bibitem{Mukhi3}
S. Mukhi and C. Vafa,
Two dimensional black-hole as topological coset model of c=1 string theory, arXiv:hep-th/9301083.
\bibitem{Susskind}
L. Susskind, Matrix theory black holes and the Gross Witten transition,
in "Trieste 1998. Nonpertubative aspects of strings, branes and supersymmetry", 
P. 390 (1998), World Scientific, Singapore.
\bibitem{Wadia}
L.Alvarez-Gaume, P.Basu, M, Marino and S.R.Wadia, Blackhole/String Transition for the small 
Schwarzschild Blackhole of $AdS_5 \times S^5$ and 
Critical Unitary Matrix Models, arXiv:hep-th/0605041.
\bibitem{Spohn}
   M. Pr\"ahofer and H. Spohn, 
Scale invariance of the PNG droplet and the Airy process,
J. Stat. Phys. 108 (2002) 1071.
\bibitem{Bleher}
   A. Aptekarev, P. Bleher and A. Kuijlaars, Large n limit of Gaussian random matrices with external source Part II.
Commun. Math. Phys. 259 (2005) 367.
\bibitem{Tracy}
    C. Tracy and H. Widom,  The Pearcey Process, Commun. Math. Phys. 263 (2006) 381.
\bibitem{Okounkov2}
   A. Okounkov and N. Reshetikhin, Random skew plane partitions and the Pearcey process, 
Commun. Math. Phys. 269 (2007) 571.
\bibitem{BH8}
    E. Br\'ezin and S. Hikami, Universal singularity at the closure of a gap in a random matrix theory,
Phys. Rev. B 57 (1998) 4140. arXiv:cond-mat/9804023.
\bibitem{BH9}
    E. Br\'ezin and S. Hikami, Level spacing of random matrices in an external source,
     Phys. Rev. E58 (1998) 7176. arXiv:cond-mat/9804024.
\bibitem{BH4aa}
   E. Br\'ezin and S. Hikami, Intersection numbers from antisymmetric Gaussian  matrix model,
     JHEP07(2009)050. arXiv:0804.4531
\bibitem{BH5}
E. Br\'ezin and S. Hikami, Extension of 
level-spacing universality. Phys. Rev. E56, 264 (1997). arXiv:cond-mat/9702213.
\bibitem{BH6}
E. Br\'ezin and S. Hikami, 
Spectral form factor in a random matrix theory, Phys. Rev. E55, 4067 (1997). arXiv:cond-mat/9608116.
\bibitem{Gross}
  D. Gross and Newman, Unitary and Hermitian Matrices In An External Field II:
The Kontsevich Model And Continuum Virasoro Constraints. arXiv: hep-th/9112069.
\bibitem{Adler}
M. Adler and P. van Moerbeke, A matrix integral solution to two-dimensional $W_p$-gravity,
Commun. Math. Phys. 147 (1992) 25.
\bibitem{Okounkov}
  A. Okounkov, Generating functions for intersection numbers on moduli spaces of curves,
Int. Math. Res. Not. 18, 933 (2002).
\bibitem{Dijkgraf}
   R. Dijkgraaf, Intersection theory, integrable hierarchies and 
topological field theory, IASSNS-HEP-91/91 (1991), 
In {\it New symmetry principles in quantum field theory (Carg\`ese,1991)},
page 95-158 Plenum, New York,1992.
\bibitem{IZ}
  C. Itzykson and B.-J.Zuber, 
Combinatorics of the moduler group II The Kontsevich integrals.
Intern. Journ. Mod. Phys. A7 (1992) 5661. arXiv:hep-th/9201001.
\bibitem{MS}
  Y. Makeenko and G.W.Semenoff,
   Properties of hermitian matrix models in an external field,
   Mod. Phys. Lett. A6 (1991) 3455.
\bibitem{BH4aaa}
E. Br\'ezin and S. Hikami, New correlation functions for random matrices and integrals over
supergroups. J. Phys. A 36 (2003) 711.
\bibitem{HZ}
    J.Harer and D.Zagier, 
The Euler characteristic of the moduli space of curves, 
Invent. Math. 85 (1986) 457.
\bibitem{Penner}
    R.C.Penner, Perturbative series and the moduli space of Riemann surfaces, 
J. Diff. Geometry, 27 (1988) 35.
\bibitem{BH7}
    E. Br\'ezin and S. Hikami, Characteristic Polynomials of Random Matrices, 
Commun. Math. Phys. 214 (2000) 111. arXiv:math-ph/9910005.
\bibitem{Mukhi3}
   A. Mukherjee and S. Mukhi, c=1 matrix models: equivalences and open-closed string duality,
   arXiv:hep-th/0505180. and refereces there in.
\end{thebibliography}
\end{document}